\documentclass[twocolumn,pra,aps,showpacs,superscriptaddress,floatfix,showkeys,nofootinbib,10pt]{revtex4-2}
\usepackage{CJK}
\usepackage{amsfonts}
\usepackage{graphicx}
\usepackage{amssymb}
\usepackage{amsmath}
\usepackage{natbib}

\newcommand{\ket}[1]{\left | #1 \right \rangle}
\newcommand{\bra}[1]{\left \langle #1 \right |}

\begin{document}
\title{Multipartite quantum cryptography based on the violation of Svetlichny's inequality}

\author{Yang Xiang}
\email{xiangyang@vip.henu.edu.cn (corresponding author)}
\affiliation{School of Physics and Electronics, Henan University, Kaifeng, Henan 475004, China}

\date{\today}
\begin{abstract}
Multipartite cryptography is useful for some particular missions. In this paper, we present a quantum key distribution scheme in which three separated observers can securely share a set of keys by using a sequence of $3$-particle GHZ states. We prove that the violation of Svetlichny's inequality can be utilized to test for eavesdropping, and even when the eavesdropper can completely control the outcomes of two participants' measurements, our scheme still ensures the security of the keys distribution. This scheme can be easily extended to the case of $N$-party keys distribution, and the violation of $N$-partite Svetlichny's inequality guarantees the security of the generalized scheme.
Since the GHZ state has maximum entanglement, its perfect monogamy guarantee the device-independent security of our protocol.
However quantum entanglement is a vulnerable resource which is often decayed during transmission, so we need here to derive the secret-key rate of our protocol under the condition of using quantum states with non-maximal entanglement.
We then calculate the extractable secret-key rate of the three-party key distribution protocol for the Werner state in the device-independent scenario. We find that the value of the extractable secret-key rate monotonously approaches $1$ as the value of the visibility of the Werner state increases, and it reaches its maximum value $1$ when the Werner state becomes the GHZ state.

\end{abstract}

\maketitle




\section{Introduction}

The simplest private keys are two identical or opposite strings of random bits, which are shared by sender and receiver and can be employed to secure communication between them. The security of communication is based entirely on the privacy of keys, while the latter is built on the secure distribution of it. Quantum cryptography or quantum key distribution (QKD) is a procedure which exploits quantum principles to secure the distribution of keys. In $1984$, Bennett and Brassard \cite{bennett1984quantum} presented the first protocol of QKD (BB$84$). In $1991$, Ekert \cite{PhysRevLett.67.661} proposed a scheme of QKD in which Bell's inequality \cite{bell1964einstein,bell1988speakable,PhysRevLett.23.880} had been used to test for eavesdropping. In the same year,  Bennett, Brassard and Mermin \cite{bennett1992quantum} proposed a simpler but conceptually equivalent version of Ekert's scheme, in which they use Einstein-Podolsky-Rosen correlations \cite{PhysRev.47.777} both to construct key strings and to test for eavesdropping. Since then numerous QKD protocols have been presented \cite{PhysRevLett.68.557,RevModPhys.74.145,gisin1997quantum,PhysRevLett.87.117901,PhysRevA.65.012311,PhysRevLett.86.1911,PhysRevLett.87.010403,
PhysRevLett.90.160408,beige2002secure,PhysRevA.67.012311},
quantum cryptography is one of the fastest growing areas in quantum information science\cite{PhysRevLett.94.230504,RevModPhys.81.1301,PhysRevA.93.032338,
PhysRevA.93.052303,PhysRevA.95.010101,PhysRevLett.118.220501,PhysRevA.100.042329}
and many research results have been applied to commercial applications\cite{PhysRevLett.119.200501,pirandola2020advances,PRXQuantum.2.010304,PRXQuantum.3.020341}.

\begin{figure}[t]
\includegraphics[width=0.95\columnwidth,
height=0.65\columnwidth]{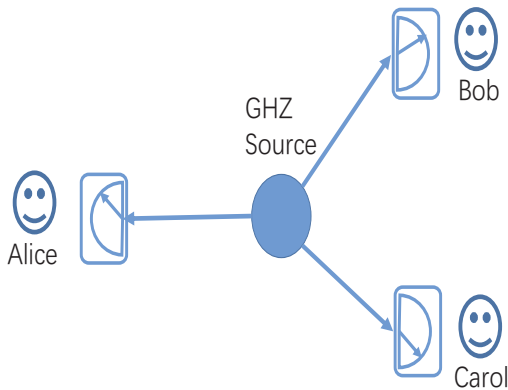} \caption{There exist a particle source which can generate a sequence of $3$-particle in GHZ states $\ket{\Psi}_{GHZ}=\frac{1}{2}\left(\ket{\uparrow\uparrow\uparrow}+\ket{\downarrow\downarrow\downarrow}\right)$ and then these particles are sent to Alice, Bob, and Carol respectively. Upon receiving their particles, Alice, Bob, and Carol can randomly choose measurements from their own sets of alternative measurement settings respectively. After all particles have been measured, Alice, Bob, and Carol announce their measurement choices publicly. In all these measurement setting options, about $1/4$ of them can be used to construct tripartite keys, about $1/2$ of them can be used to calculate whether the SI inequality will be violated, and only $1/4$ of them are useless.}
\label{fig1}
\end{figure}

\begin{figure}[t]
\includegraphics[width=0.95\columnwidth,
height=0.65\columnwidth]{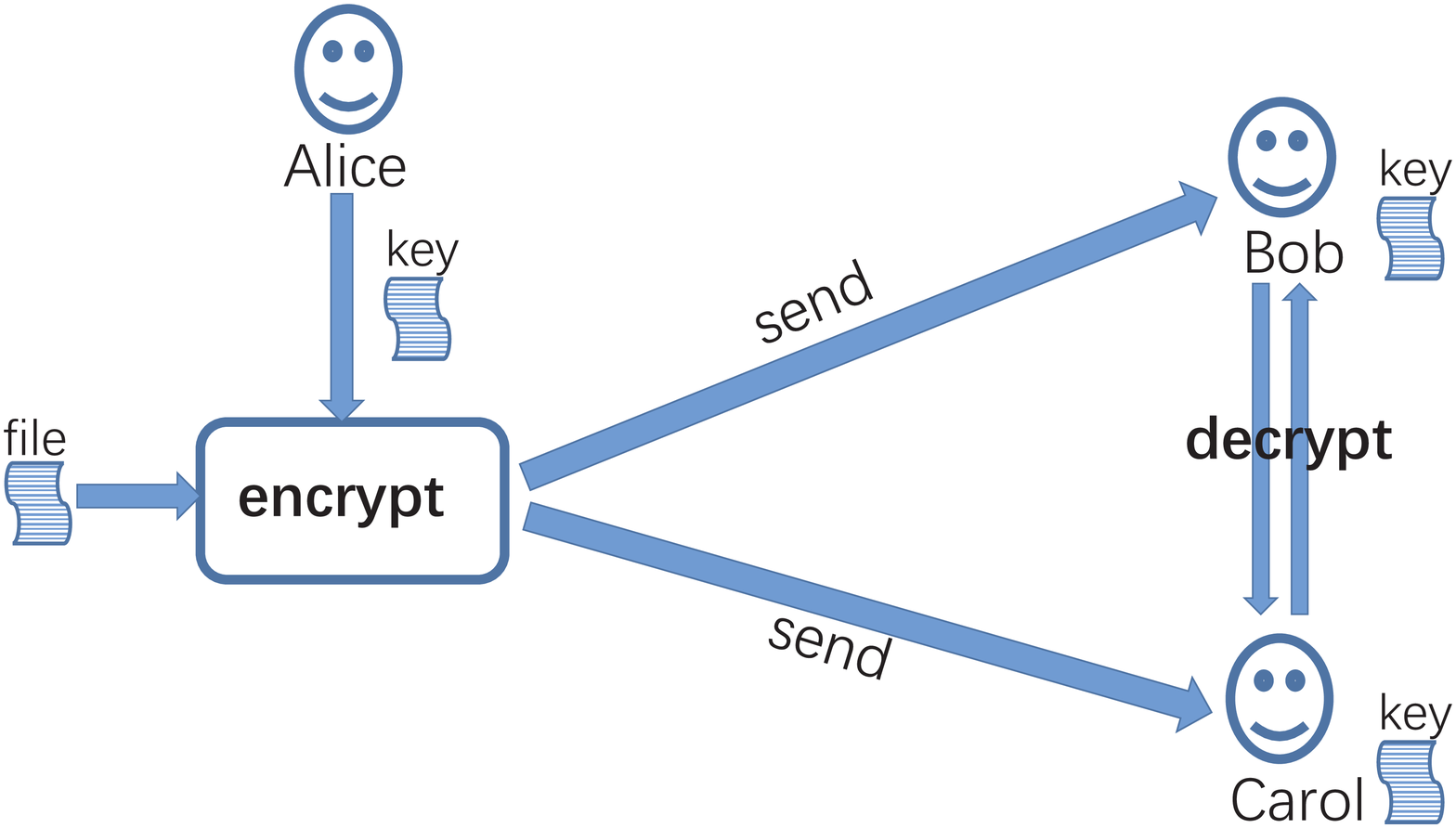} \caption{Alice has a file to pass to Bob and Carol, she requires that Bob and Carol can only open this file together, and neither of them can open the file alone. If these three people have shared tripartite private keys in advance, Alice can encrypt this file by using her keys and then send it to Bob and Carol. Since for tripartite private keys, no one can infer the keys of either of the other two from the keys in their own hands, Bob and Carol definitely can't decrypt the file alone. So Bob and Carol must cooperate to decrypt the file.}
\label{fig2}
\end{figure}

In addition to the fully studied two-party keys, there are private keys that involve multipartite communications. Analogous to bipartite private keys, multipartite private keys consist of strings of random bits that shared by multiple parties, but in this case these strings have collective correlations rather than pairwise correlations (Fig.\ref{fig1}). Multipartite private keys can apply to some tasks that bipartite private keys cannot accomplish. For example, the situation in Fig.\ref{fig2}, Alice has a file to pass to Bob and Carol, she requires that Bob and Carol can only open this file together, and neither of them can open the file alone. If these three people have shared tripartite private keys in advance, Alice can encrypt this file by using her keys and then send it to Bob and Carol. Since for the case of tripartite private keys, no one can infer the keys of either of the other two from the keys in their own hands, Bob and Carol definitely can't decrypt the file alone. We can also design a lot of tasks in which multi-party keys are applicable but two-party keys are not competent.

In the secure distributions of two-party keys, one use the bipartite quantum correlations, while in the secure distributions of multi-party keys, as the illustration of Fig.\ref{fig1} we need to use the genuine multipartite quantum correlations (GMQC)\cite{PhysRevLett.106.020405,PhysRevLett.106.250404,PhysRevLett.88.210401,PhysRevLett.107.210403,PhysRevA.89.032117,RevModPhys.86.419}. GMQC is a collective correlation which involve all subsystems, so it cannot be reduced to mixtures of states in which a smaller number of subsystems are entangled. There exist many inequivalent types of GMQC\cite{xiang2011bound}, and its structure is much richer than that of bipartite correlations\cite{PhysRevA.71.022101,pironio2011extremal}. Svetlichny proposed the first method to detect GMQC, Svetlichny's inequality (SI) \cite{PhysRevD.35.3066,PhysRevLett.89.060401} is a Bell-like inequality, the violation of which can be used to confirm the existence of GMQC. GHZ states is a quantum state with GMQC\cite{greenberger1990bell,pan2000experimental}, by using it and proper measurement settings one can result in maximal violation of SI \cite{PhysRevLett.89.060401}.

There has been a lot of theoretical and experimental works on multi-party key distribution. For example, in \cite{Grasselli2018} the authors
introduced an $N$-party version of the BB$84$ protocol, and in \cite{Grasselli2019} the same authors introduced a new multi-party
QKD protocol that exploits $N$-partite W state to establish a secret multi-party key among the $N$ users. In \cite{PhysRevLett.114.090501} and \cite{sciadv.abe0395}, some experimentally feasible schemes have been proposed which manifest the possibility for practical realization of multi-party key distribution over long-distance. In addition, there are already some multi-party QKD protocols whose security can be guaranteed by violations of different quantum nonlocality inequalities\cite{PhysRevLett.108.100402,PhysRevA.97.022307,PhysRevResearch.2.023251}.

In this paper, we present a scheme for secure distributions of multipartite keys by using GHZ states (Fig.\ref{fig1}), in this scheme the violation of SI can be used to test for eavesdropping. We first discuss tripartite keys. In this case after receiving their particles, Alice, Bob, and Carol can randomly choose measurements from their own sets of alternative measurement settings respectively. After all particles have been measured, Alice, Bob, and Carol announce their measurement choices publicly. In all measurement setting options, the outcomes of about $1/4$ of them can be used to construct tripartite keys, the outcomes of about $1/2$ of them can be used to calculate whether the SI inequality will be violated, and only the outcomes of about $1/4$ of them are useless. We then extend this scheme to the case of $N$-party keys. In this case we find that the violation of N-partite SI still guarantees the security of the generalized scheme, and also only a quarter of the measurement options are useless. In particular, we prove that in an extreme case, where the eavesdropper can completely control $N-1$-party's measurement outcomes, the violation of SI still ensures the security of the keys distribution.
Since GHZ states is a maximum entanglement state, whose perfect monogamy guarantee the device-independent security of our protocol.
However quantum entanglement is a vulnerable resource which is often decayed during transmission, so we need here to derive the secret-key rate of our protocol under the condition of using quantum states with non-maximal entanglement. We then calculate the extractable secret-key rate of our three-party QKD protocol for the Werner state in the device-independent scenario. We find that the value of the extractable secret-key rate monotonously approaches $1$ as the value of the visibility of the Werner state increases, and it reaches its maximum value $1$ when the Werner state becomes the GHZ state.


\section{Three-party key distribution}

Our scheme uses a series of 3-particle GHZ states. We can assume there exist a particle source which can generate a sequence of $3$-particle in GHZ states $\ket{\Psi}_{GHZ}=\frac{1}{\sqrt{2}}\left(\ket{\uparrow\uparrow\uparrow}+\ket{\downarrow\downarrow\downarrow}\right)$ and then these particles are sent to Alice, Bob, and Carol respectively.
After receive his (her) particle, Alice, Bob, and Carol can perform spin measurements on their own particles along some directions, we use unit vectors $\mathbf{a}_{i}$, $\mathbf{b}_{j}$, and $\mathbf{c}_{k}$ to stand for their directions of spin measurements. For simplicity, we require all $\mathbf{a}_{i}$, $\mathbf{b}_{j}$, and $\mathbf{c}_{k}$ lie in the $x-y$ plane, and all vectors are characterized by their azimuthal angles: $\alpha_{i}$, $\beta_{j}$, and $\gamma_{k}$. We also use these azimuthal angles to represent their spin measurements, for example Alice's spin measurements are $\hat{A}_{i}=\mathbf{a}_{i}\cdot\mathbf{\sigma}=\cos{\alpha_{i}}\sigma_{x}+\sin{\alpha_{i}}\sigma_{y}$, and similarly $\hat{B}_{j}=\cos{\beta_{j}}\sigma_{x}+\sin{\beta_{j}}\sigma_{y}$ \text($\hat{C}_{k}=\cos{\gamma_{k}}\sigma_{x}+\sin{\gamma_{k}}\sigma_{y}$\text) for Bob's  (Carol's). Now we assume that for every particle she received Alice randomly choose a measurement from \{$\alpha_{0}=-\frac{\pi}{4}, \alpha_{1}=\frac{\pi}{4}, \alpha_{2}=0,\alpha_{3}=\frac{\pi}{2}$\}, Bob randomly choose a measurement from \{${\beta_{0}=0, \beta_{1}=\frac{\pi}{2}}$\}, and similarly Carol randomly choose a measurement from \{${\gamma_{0}=0, \gamma_{1}=\frac{\pi}{2}}$\}. It's obvious that there are a total of $16$ measurement options for a $3$-particle GHZ state, we will see that outcomes of four measurement options can be used to construct three-party private keys. From $\sigma_{x}\ket{\uparrow}=\ket{\downarrow}$, $\sigma_{x}\ket{\downarrow}=\ket{\uparrow}$, $\sigma_{y}\ket{\uparrow}=i\ket{\downarrow}$, and $\sigma_{y}\ket{\downarrow}=-i\ket{\uparrow}$, we can get the following equations,
\begin{eqnarray}
&&(i)~~~~~\sigma_{x}\otimes\sigma_{y}\otimes\sigma_{y}\ket{\Psi}_{GHZ}=-\ket{\Psi}_{GHZ}\nonumber\\
&&(ii)~~~~\sigma_{y}\otimes\sigma_{x}\otimes\sigma_{y}\ket{\Psi}_{GHZ}=-\ket{\Psi}_{GHZ}\nonumber\\
&&(iii)~~~\sigma_{y}\otimes\sigma_{y}\otimes\sigma_{x}\ket{\Psi}_{GHZ}=-\ket{\Psi}_{GHZ}\nonumber\\
&&(iv)~~~~\sigma_{x}\otimes\sigma_{x}\otimes\sigma_{x}\ket{\Psi}_{GHZ}=\ket{\Psi}_{GHZ}.
\label{keycalculate}
\end{eqnarray}
From above equations, we find that for some certain measurement settings their measurement outcomes have definite collective correlations. So for these four measurement settings of $(0,0,0)$, $(0,\frac{\pi}{2},\frac{\pi}{2})$, $(\frac{\pi}{2},0,\frac{\pi}{2})$, and $(\frac{\pi}{2},\frac{\pi}{2},0)$, their outcomes can be used as tripartite keys.

We briefly introduce SI which will be used to test for eavesdropping. Tripartite SI can expressed as the following equation,
\begin{eqnarray}
&&\big|\langle A_{0}B_{0}C_{0}\rangle+\langle A_{0}B_{0}C_{1}\rangle+\langle A_{0}B_{1}C_{0}\rangle+\langle A_{1}B_{0}C_{0}\rangle\nonumber\\
&&-\langle A_{0}B_{1}C_{1}\rangle-\langle A_{1}B_{0}C_{1}\rangle-\langle A_{1}B_{1}C_{0}\rangle-\langle A_{1}B_{1}C_{1}\rangle\big|\nonumber\\
&&\leq 4.
\label{trisi}
\end{eqnarray}
Where all $\langle A_{i}B_{j}C_{k}\rangle$'s represent average values of $ A_{i}B_{j}C_{k}$'s, $A_{0}$ and $A_{1}$ are Alice's outcomes for corresponding measurements $\hat{A}_{0}$ and $\hat{A}_{1}$, and similarly $B_{0}$ and $B_{1}$ ($C_{0}$ and $C_{1}$) for Bob's (Carol's). As mentioned above, we use unit vector $\mathbf{a}_{i}$, $\mathbf{b}_{j}$, and $\mathbf{c}_{k}$ to stand for directions of spin measurements $\hat{A}_{i}$, $\hat{B}_{j}$, and $\hat{C}_{k}$ respectively, and restrict them lie in the $x-y$ plane. As above we use their azimuthal angles $\alpha_{i}$, $\beta_{j}$, and $\gamma_{k}$ to represent these measurements $\hat{A}_{i}$, $\hat{B}_{j}$, and $\hat{C}_{k}$ respectively.
For $3$-particle in GHZ state of $\ket{\Psi}_{GHZ}=\frac{1}{\sqrt{2}}\left(\ket{\uparrow\uparrow\uparrow}+\ket{\downarrow\downarrow\downarrow}\right)$, we choose the measurement protocol \cite{PhysRevLett.89.060401}
\begin{eqnarray}
\left(\alpha_{0},\beta_{0},\gamma_{0}\right)=\left(-\frac{\pi}{4},0,0\right),\nonumber\\
\left(\alpha_{1},\beta_{1},\gamma_{1}\right)=\left(\frac{\pi}{4},\frac{\pi}{2},\frac{\pi}{2}\right),
\label{protocal3}
\end{eqnarray}
then we can achieve the maximal value $4\sqrt{2}$ of Eq. (\ref{trisi}),
i.e. the maximal violation of SI \cite{PhysRevLett.89.060401}.
So we find that in above all $16$ measurement options there are these eight measurement settings of $(\pm\frac{\pi}{4},0,0)$, $(\pm\frac{\pi}{4},0,\frac{\pi}{2})$, $(\pm\frac{\pi}{4},\frac{\pi}{2},0)$, and $(\pm\frac{\pi}{4},\frac{\pi}{2},\frac{\pi}{2})$, whose outcomes can be used to calculate the value of SI.

Therefore, our scheme is as follows:\\
$(1)$. There is a particle source which can generate a sequence of $3$-particle in GHZ states and then send these particles to Alice, Bob, and Carol respectively.\\
$(2)$. After receiving their particles, for every particle Alice randomly choose to perform a measurement from  \{${\alpha_{0}=-\frac{\pi}{4}, \alpha_{1}=\frac{\pi}{4}, \alpha_{2}=0,\alpha_{3}=\frac{\pi}{2}}$\}, Bob randomly choose a measurement from \{${\beta_{0}=0, \beta_{1}=\frac{\pi}{2}}$\}, and similarly Carol randomly choose a measurement from \{${\gamma_{0}=0, \gamma_{1}=\frac{\pi}{2}}$\}. After all particles have been measured, Alice, Bob, and Carol announce their all measurement choices publicly.\\
$(3)$. They reveal publicly the outcomes of the eight measurement settings of $(\pm\frac{\pi}{4},0,0)$, $(\pm\frac{\pi}{4},0,\frac{\pi}{2})$, $(\pm\frac{\pi}{4},\frac{\pi}{2},0)$, and $(\pm\frac{\pi}{4},\frac{\pi}{2},\frac{\pi}{2})$ that involving calculation of the value of SI.They use the outcomes of the four measurement settings of $(0,0,0)$, $(0,\frac{\pi}{2},\frac{\pi}{2})$, $(\frac{\pi}{2},0,\frac{\pi}{2})$, and $(\frac{\pi}{2},\frac{\pi}{2},0)$ as tripartite keys.\\
$(4)$. If they find the violation of SI, they can ensure the security of the key distribution.

Below we prove that the violation of SI can be used to test for eavesdropping. We first consider a mild case in which the eavesdropper disturb the particles in their transmissions from the source to the legitimate users. And then we'll consider an extreme situation. If the eavesdropper disturb (measure) the particles in their transmissions, their quantum states will collapse to the eigenstates of spin in some certain directions. We use unit vectors $\mathbf{n}_{a}$, $\mathbf{n}_{b}$, $\mathbf{n}_{c}$ to denote the directions of the collapsed spin eigenstates of particles a, b, and c that have been sent to Alice, Bob, and Carol respectively. The normalized probability measure $\rho(\mathbf{n}_{a}, \mathbf{n}_{b}, \mathbf{n}_{c})$ exactly reflects the eavesdropper's measurement strategy, and it also determines the correlations of the measurement results of the legitimate users', i.e. $\langle A_{i}B_{j}C_{k}\rangle=\int\rho(\mathbf{n}_{a}, \mathbf{n}_{b}, \mathbf{n}_{c})d\mathbf{n}_{a} d\mathbf{n}_{b} d\mathbf{n}_{c}(\mathbf{a}_{i}\cdot\mathbf{n}_{a})(\mathbf{b}_{j}\cdot\mathbf{n}_{b})(\mathbf{c}_{k}\cdot\mathbf{n}_{c})$. Since all $\mathbf{a}_{i}$, $\mathbf{b}_{j}$, and $\mathbf{c}_{k}$ lie in the $x-y$ plane, it's obvious
that all unit vectors $\mathbf{n}_{a}$, $\mathbf{n}_{b}$, $\mathbf{n}_{c}$ should also lie in the $x-y$ plane in order to increase the possibility of violation of SI, i.e. the eavesdropper should measure particles along the directions in the $x-y$ plane.
We find that the left-hand side of Eq. (\ref{trisi}) can be expressed as
\begin{eqnarray}
&&\big|\langle( A_{0}B_{0}+A_{0}B_{1}+A_{1}B_{0}-A_{1}B_{1})C_{0}\nonumber\\
&&-(A_{0}B_{1}+A_{1}B_{0}+A_{1}B_{1}-A_{0}B_{0})C_{1}\rangle\big|\nonumber\\
&&\leq \big|\langle( A_{0}B_{0}+A_{0}B_{1}+A_{1}B_{0}-A_{1}B_{1})C_{0}\rangle\big|\nonumber\\
&&+\big| \langle(A_{0}B_{1}+A_{1}B_{0}+A_{1}B_{1}-A_{0}B_{0})C_{1}\rangle\big|.
\label{si1}
\end{eqnarray}
We now calculate both terms on the right-hand side of the  above inequality, and prove that the sum of them is less than $2$. The first term
\begin{eqnarray}
&&\big|\langle( A_{0}B_{0}+A_{0}B_{1}+A_{1}B_{0}-A_{1}B_{1})C_{0}\rangle\big|\nonumber\\
&&=\Big|\int \rho(\mathbf{n}_{a}, \mathbf{n}_{b}, \mathbf{n}_{c}) d\mathbf{n}_{a} d\mathbf{n}_{b} d\mathbf{n}_{c}[(\mathbf{a}_{0}\cdot\mathbf{n}_{a})(\mathbf{b}_{0}\cdot\mathbf{n}_{b})\nonumber\\
&&+(\mathbf{a}_{0}\cdot\mathbf{n}_{a})(\mathbf{b}_{1}\cdot\mathbf{n}_{b})
+(\mathbf{a}_{1}\cdot\mathbf{n}_{a})(\mathbf{b}_{0}\cdot\mathbf{n}_{b})\nonumber\\
&&-(\mathbf{a}_{1}\cdot\mathbf{n}_{a})(\mathbf{b}_{1}\cdot\mathbf{n}_{b})](\mathbf{c}_{0}\cdot\mathbf{n}_{c})\Big|\nonumber\\
&&\leq \int\rho'(\mathbf{n}_{a}, \mathbf{n}_{b}) d\mathbf{n}_{a} d\mathbf{n}_{b}\Big|[(\mathbf{a}_{0}\cdot\mathbf{n}_{a})(\mathbf{b}_{0}\cdot\mathbf{n}_{b})\nonumber\\
&&+(\mathbf{a}_{0}\cdot\mathbf{n}_{a})(\mathbf{b}_{1}\cdot\mathbf{n}_{b})
+(\mathbf{a}_{1}\cdot\mathbf{n}_{a})(\mathbf{b}_{0}\cdot\mathbf{n}_{b})\nonumber\\
&&-(\mathbf{a}_{1}\cdot\mathbf{n}_{a})(\mathbf{b}_{1}\cdot\mathbf{n}_{b})]\Big|
\label{si2}.
\end{eqnarray}
Where $\mathbf{a}_{0}, \mathbf{b}_{0}, \mathbf{c}_{0}$ ($\mathbf{a}_{1}, \mathbf{b}_{1}, \mathbf{c}_{1}$) stand for the directions of spin measurements, they all lie in the $x-y$ plane and their azimuthal
angles can be see in  Eq. (\ref{protocal3}). The normalized probability measure $\rho'(\mathbf{n}_{a}, \mathbf{n}_{b})=\int\rho(\mathbf{n}_{a}, \mathbf{n}_{b}, \mathbf{n}_{c})d\mathbf{n}_{c}$. We calculate the integrand in Eq. (\ref{si2}),
\begin{eqnarray}
&&\Big|[(\mathbf{a}_{0}\cdot\mathbf{n}_{a})(\mathbf{b}_{0}\cdot\mathbf{n}_{b})+(\mathbf{a}_{0}\cdot\mathbf{n}_{a})(\mathbf{b}_{1}\cdot\mathbf{n}_{b})\nonumber\\
&&+(\mathbf{a}_{1}\cdot\mathbf{n}_{a})(\mathbf{b}_{0}\cdot\mathbf{n}_{b})-(\mathbf{a}_{1}\cdot\mathbf{n}_{a})(\mathbf{b}_{1}\cdot\mathbf{n}_{b})]\Big|\nonumber\\
&&=\Big|(\mathbf{a}_{0}\cdot\mathbf{n}_{a})[(\mathbf{b}_{0}+\mathbf{b}_{1})\cdot\mathbf{n}_{b}]
+(\mathbf{a}_{1}\cdot\mathbf{n}_{a})[(\mathbf{b}_{0}-\mathbf{b}_{1})\cdot\mathbf{n}_{b}]\Big|\nonumber\\
&&=\bigg|(\mathbf{a}_{0}\cdot\mathbf{n}_{a})\bigg[(\hat{x}+\hat{y})\cdot\mathbf{n}_{b}\bigg]
+(\mathbf{a}_{1}\cdot\mathbf{n}_{a})\bigg[(\hat{x}-\hat{y})\cdot\mathbf{n}_{b}\bigg]\bigg|\nonumber\\
&&=\sqrt{2}\bigg|\bigg[\frac{\hat{x}-\hat{y}}{\sqrt{2}}\cdot\mathbf{n}_{a}\bigg]\bigg[\frac{\hat{x}+\hat{y}}{\sqrt{2}}\cdot\mathbf{n}_{b}\bigg]\nonumber\\
&&+\bigg[\frac{\hat{x}+\hat{y}}{\sqrt{2}}\cdot\mathbf{n}_{a}\bigg]\bigg[\frac{\hat{x}-\hat{y}}{\sqrt{2}}\cdot\mathbf{n}_{b}\bigg]\bigg|\nonumber\\
&&=\sqrt{2}\big|-\big[\cos{\theta_{a}}\sin{\theta_{b}}-\sin{\theta_{a}}\cos{\theta_{b}}\big]\big|\nonumber\\
&&=\sqrt{2}\big|\sin{(\theta_{a}+\theta_{b})}\big|
\label{si3}.
\end{eqnarray}
Where $\hat{x}$ and $\hat{y}$ represent the coordinate directions of the x-axis and the y-axis, $\theta_{a}$ and $\theta_{b}$ are the angles between the vector $\frac{\hat{x}-\hat{y}}{\sqrt{2}}$ and
$\mathbf{n}_{a}$ and $\mathbf{n}_{b}$ respectively. Substitute the result of Eq. (\ref{si3}) into Eq. (\ref{si2}), we get a upper bound of $\mid \langle( A_{0}B_{0}+A_{0}B_{1}+A_{1}B_{0}-A_{1}B_{1})C_{0}\rangle\mid$.

In a similar way we calculate $\mid \langle(A_{0}B_{1}+A_{1}B_{0}+A_{1}B_{1}-A_{0}B_{0})C_{1}\rangle\mid$.
\begin{eqnarray}
&&\left|\langle( A_{0}B_{1}+A_{1}B_{0}+A_{1}B_{1}-A_{0}B_{0})C_{1}\rangle\right|\nonumber\\
&&=\Big|\int \rho(\mathbf{n}_{a}, \mathbf{n}_{b}, \mathbf{n}_{c}) d\mathbf{n}_{a} d\mathbf{n}_{b} d\mathbf{n}_{c}[(\mathbf{a}_{0}\cdot\mathbf{n}_{a})(\mathbf{b}_{1}\cdot\mathbf{n}_{b})\nonumber\\
&&+(\mathbf{a}_{1}\cdot\mathbf{n}_{a})(\mathbf{b}_{0}\cdot\mathbf{n}_{b})
+(\mathbf{a}_{1}\cdot\mathbf{n}_{a})(\mathbf{b}_{1}\cdot\mathbf{n}_{b})\nonumber\\
&&-(\mathbf{a}_{0}\cdot\mathbf{n}_{a})(\mathbf{b}_{0}\cdot\mathbf{n}_{b})](\mathbf{c}_{1}\cdot\mathbf{n}_{c})\Big|\nonumber\\
&&\leq \int\rho'(\mathbf{n}_{a}, \mathbf{n}_{b}) d\mathbf{n}_{a} d\mathbf{n}_{b}\Big|[(\mathbf{a}_{0}\cdot\mathbf{n}_{a})(\mathbf{b}_{1}\cdot\mathbf{n}_{b})\nonumber\\
&&+(\mathbf{a}_{1}\cdot\mathbf{n}_{a})(\mathbf{b}_{0}\cdot\mathbf{n}_{b})
+(\mathbf{a}_{1}\cdot\mathbf{n}_{a})(\mathbf{b}_{1}\cdot\mathbf{n}_{b})\nonumber\\
&&-(\mathbf{a}_{0}\cdot\mathbf{n}_{a})(\mathbf{b}_{0}\cdot\mathbf{n}_{b})]\Big|
\label{si2-2},
\end{eqnarray}
as in Eq. (\ref{si2}), $\mathbf{a}_{0}, \mathbf{b}_{0}, \mathbf{c}_{0}$ ($\mathbf{a}_{1}, \mathbf{b}_{1}, \mathbf{c}_{1}$) all lie in the $x-y$ plane and their azimuthal angles can be see in  Eq. (\ref{protocal3}), they stand for the directions of spin measurements. The normalized probability measure $\rho'(\mathbf{n}_{a}, \mathbf{n}_{b})=\int\rho(\mathbf{n}_{a}, \mathbf{n}_{b}, \mathbf{n}_{c})d\mathbf{n}_{c}$.
We also calculate the integrand in Eq. (\ref{si2-2}),
\begin{eqnarray}
&&\Big|[(\mathbf{a}_{0}\cdot\mathbf{n}_{a})(\mathbf{b}_{1}\cdot\mathbf{n}_{b})+(\mathbf{a}_{1}\cdot\mathbf{n}_{a})(\mathbf{b}_{0}\cdot\mathbf{n}_{b})\nonumber\\
&&+(\mathbf{a}_{1}\cdot\mathbf{n}_{a})(\mathbf{b}_{1}\cdot\mathbf{n}_{b})-(\mathbf{a}_{0}\cdot\mathbf{n}_{a})(\mathbf{b}_{0}\cdot\mathbf{n}_{b})]\Big|\nonumber\\
&&=\Big|(\mathbf{a}_{0}\cdot\mathbf{n}_{a})[(-\mathbf{b}_{0}+\mathbf{b}_{1})\cdot\mathbf{n}_{b}]
+(\mathbf{a}_{1}\cdot\mathbf{n}_{a})[(\mathbf{b}_{0}+\mathbf{b}_{1})\cdot\mathbf{n}_{b}]\Big|\nonumber\\
&&=\bigg|(\mathbf{a}_{0}\cdot\mathbf{n}_{a})\bigg[(-\hat{x}+\hat{y})\cdot\mathbf{n}_{b}\bigg]
+(\mathbf{a}_{1}\cdot\mathbf{n}_{a})\bigg[(\hat{x}+\hat{y})\cdot\mathbf{n}_{b}\bigg]\bigg|\nonumber\\
&&=\sqrt{2}\bigg|\bigg[\frac{\hat{x}-\hat{y}}{\sqrt{2}}\cdot\mathbf{n}_{a}\bigg]\bigg[\frac{-\hat{x}+\hat{y}}{\sqrt{2}}\cdot\mathbf{n}_{b}\bigg]\nonumber\\
&&+\bigg[\frac{\hat{x}+\hat{y}}{\sqrt{2}}\cdot\mathbf{n}_{a}\bigg]\bigg[\frac{\hat{x}+\hat{y}}{\sqrt{2}}\cdot\mathbf{n}_{b}\bigg]\bigg|\nonumber\\
&&=\sqrt{2}\big|-\cos{\theta_{a}}\cos{\theta_{b}}+\sin{\theta_{a}}\sin{\theta_{b}}\big|\nonumber\\
&&=\sqrt{2}\big|\cos{(\theta_{a}+\theta_{b})}\big|
\label{si3-3}.
\end{eqnarray}
As in Eq. (\ref{si3}), $\theta_{a}$ and $\theta_{b}$ are still the angles between the vector $\frac{\hat{x}-\hat{y}}{\sqrt{2}}$ and
$\mathbf{n}_{a}$ and $\mathbf{n}_{b}$ respectively. Substitute the result of Eq. (\ref{si3-3}) into Eq. (\ref{si2-2}), we get a upper bound of
$\mid \langle(A_{0}B_{1}+A_{1}B_{0}+A_{1}B_{1}-A_{0}B_{0})C_{1}\rangle\mid$.

Based on the above calculation, we finally obtain
\begin{eqnarray}
&&\Big|\langle A_{0}B_{0}C_{0}\rangle+\langle A_{0}B_{0}C_{1}\rangle+\langle A_{0}B_{1}C_{0}\rangle+\langle A_{1}B_{0}C_{0}\rangle\nonumber\\
&&-\langle A_{0}B_{1}C_{1}\rangle-\langle A_{1}B_{0}C_{1}\rangle-\langle A_{1}B_{1}C_{0}\rangle-\langle A_{1}B_{1}C_{1}\rangle\Big|\nonumber\\
&&\leq \Big|\langle( A_{0}B_{0}+A_{0}B_{1}+A_{1}B_{0}-A_{1}B_{1})C_{0}\rangle\Big|\nonumber\\
&&+\Big| \langle(A_{0}B_{1}+A_{1}B_{0}+A_{1}B_{1}-A_{0}B_{0})C_{1}\rangle\Big|\nonumber\\
&&\leq \int\rho'(\mathbf{n}_{a}, \mathbf{n}_{b}) d\mathbf{n}_{a} d\mathbf{n}_{b}
\sqrt{2}\bigg[\big|\sin{(\theta_{a}+\theta_{b})}\big|\nonumber\\
&&+\big|\cos{(\theta_{a}+\theta_{b})}\big|\bigg]\nonumber\\
&&\leq 2 \int\rho'(\mathbf{n}_{a}, \mathbf{n}_{b}) d\mathbf{n}_{a} d\mathbf{n}_{b}
\bigg|\pm\sin{(\theta_{a}+\theta_{b}\pm\frac{\pi}{2})}\bigg|\nonumber\\
&&\leq 2 \int\rho'(\mathbf{n}_{a}, \mathbf{n}_{b}) d\mathbf{n}_{a} d\mathbf{n}_{b}=2.
\label{si4}
\end{eqnarray}
We can see that there is no violation of SI in this case. So if the eavesdropper disturb the particles, the legitimate users always find that key distribution fails.

Now we consider an extreme situation in that the eavesdropper has the ability to control the measurement results of two participants, we want to know whether the eavesdropper is thus capable of creating the illusion of successful key distribution, i.e. a violation of SI. Without loss of generality we assume that the eavesdropper can completely control the measurement results of Alice and Bob. We notice that these two functions
$( A_{0}B_{0}+A_{0}B_{1}+A_{1}B_{0}-A_{1}B_{1})$ and $(A_{0}B_{1}+A_{1}B_{0}+A_{1}B_{1}-A_{0}B_{0})$ are not independent since they consist of the same four quantities, whenever one of the two functions reaches its maximum absolute value $4$, the other one will be $0$. If we change one quantity of all $ A_{i}B_{j}$'s to its opposite number, the value of the function with maximum absolute value $4$ will become $\pm2$, and the value of the other function will become $\pm2$ too. The discussion shows that in no case can the sum of the absolute values of these two functions exceed $4$.
So even if the eavesdropper can control the measurement results of two participants, the right-hand side of Eq. (\ref{si1}) cannot be greater than $4$.

\section{$N$-party key distribution}

We will use a series of $N$-particle GHZ state of $\ket{\Psi}^{N}_{GHZ}=\frac{1}{\sqrt{2}}\left(\ket{\uparrow}^{\otimes N}_{z}+\ket{\downarrow}^{\otimes N}_{z}\right)$ in the distribution of $N$-party key. We use $A^{(i)}_{x_i}$ to stand for the outcomes of the measurement operator $\hat{A}^{(i)}_{x_i}$ of the i-th participant, where ${x_i}$ represents the measurement choices of the i-th participant. In $N$-partite SI each person has two measurement options, so every ${x_i}$ takes the value $0$ or $1$. Note, however, that one participant in our protocol of the $N$-party key distribution had four measurement options. As in the three-party case, we use unit vectors $\mathbf{a}^{(i)}_{x_i}$ to represent the directions of spin measurements $\hat{A}^{(i)}_{x_i}$. For simplicity, we require all $\mathbf{a}^{(i)}_{x_i}$'s lie in the $x-y$ plane, and all vectors are characterized by their azimuthal angles $\alpha^{(i)}_{x_i}$. We can use these azimuthal angles to represent these measurements $\hat{A}^{(i)}_{x_i}=\mathbf{a}^{(i)}_{x_i}\cdot\mathbf{\sigma}=\cos{\alpha^{(i)}_{x_i}}\sigma_{x}+\sin{\alpha^{(i)}_{x_i}}\sigma_{y}$.

\emph{$N$-partite SI}. We can express $N$-partite SI as
\begin{eqnarray}
\left|\langle S^{\pm}_{N}\rangle\right|&=&\left|\langle\sum_{\{x_{i}\}} {v^{\pm}_{k}A^{(1)}_{x_1}A^{(2)}_{x_2}\cdot\cdot\cdot A^{(N)}_{x_N}}\rangle\right|\nonumber\\
&\leq& 2^{N-1},
\label{nsi}
\end{eqnarray}
where $S^{\pm}_{N}$ is the $N$-partite Svetlichny's operator, $\{x_{i}\}$ stands for an $N$-tuple $x_{1},...,x_{N}$, the sum is over all these tuples. The $v^{\pm}_{k}$ is the sign function of the corresponding term $A^{(1)}_{x_1}A^{(2)}_{x_2}\cdot\cdot\cdot A^{(N)}_{x_N}$, it is given by $v^{\pm}_{k}=(-1)^{[k(k\pm1)/2]}$, where $k$ is the number of times index $1$ appears in $(x_{1},x_{2},...,x_{N})$.
For $N$-particle GHZ state $\frac{1}{\sqrt{2}}\left(\ket{\uparrow}^{\otimes N}_{z}+\ket{\downarrow}^{\otimes N}_{z}\right)$, by using the following measurement protocol \cite{PhysRevLett.89.060401} one can achieve the maximal violation $2^{N-1}\sqrt{2}$ of Eq. (\ref{nsi})
\begin{eqnarray}
&&\left(\alpha^{(1)}_{0},\alpha^{(2)}_{0},...,\alpha^{(N)}_{0}\right)=\left(\frac{\pi}{4},0,...,0\right)\nonumber\\
&&\left(\alpha^{(1)}_{1},\alpha^{(2)}_{1},...,\alpha^{(N)}_{1}\right)=\left(\frac{3\pi}{4},\frac{\pi}{2},...,\frac{\pi}{2}\right).
\label{protocaln}
\end{eqnarray}

\emph{The scheme of $N$-party key distribution}. There are $N$ participants in this scheme, in every trial these participants randomly choose a measurement from their own measurement option sets respectively, we use the azimuthal angles $\alpha^{(i)}_{x_i}$ to express these measurement options and list them below
\begin{eqnarray}
&&\left(\alpha^{(1)}_{0},\alpha^{(1)}_{1},\alpha^{(1)}_{2},\alpha^{(1)}_{3}\right)=\left(\frac{\pi}{4},\frac{3\pi}{4},0,\frac{\pi}{2}\right)\nonumber\\
&&\left(\alpha^{(i)}_{0},\alpha^{(i)}_{1}\right)=\left(0,\frac{\pi}{2}\right)~~for~~i=2,3...N.
\label{scheme}
\end{eqnarray}
The first participant have four measurement options and other participants have only two, there are a total of $4\times2^{N-1}$ measurement options for a $N$-particle GHZ state.
For every trial, when all participants only measure along $\hat{x}$ or $\hat{y}$ direction and the number of the participants who measure along the $\hat{y}$ direction is even, the following equation holds
\begin{eqnarray}
&&\sigma^{(1)}_{k_{1}}\otimes\sigma^{(2)}_{k_{2}}\otimes...\otimes\sigma^{(N)}_{k_{N}}\ket{\Psi}^{N}_{GHZ}=\mp\ket{\Psi}^{N}_{GHZ}.
\label{keyn}
\end{eqnarray}
Where $\sigma^{(i)}_{k_{i}}$ is pauli operator $\sigma^{(i)}_{x}$ or $\sigma^{(i)}_{y}$, when the number of index $y$ appears in $(k_{1},k_{2},...,k_{N})$ is a multiple of $4$ the right-hand side of Eq (\ref{keyn}) takes
$\ket{\Psi}^{N}_{GHZ}$. So we can use the outcomes of these measurement options as $N$-party private keys, the number of these measurement options is $2^{N-1}$. The number of the measurement options that involved the calculation of $N$-partite SI (Eq. (\ref{protocaln})) is $2^N$, the outcomes of the remaining $2^{N-1}$ measurement options are useless.

\emph{Test for eavesdropping}. As in three-party case, we first consider a mild case where the eavesdropper measures the particles in their transmissions.
We notice that the $N$-partite Svetlichny's operator $S^{\pm}_{N}$ can be expressed as
\begin{eqnarray}
&&S^{\pm}_{N}=S^{\pm}_{N-1}A^{(N)}_{0}\mp S^{\mp}_{N-1}A^{(N)}_{1}.
\label{sio}
\end{eqnarray}
By using the result of three-party case and the mathematical induction, it is obvious that $\left|\langle S^{\pm}_{N}\rangle\right|\leq 2^{N-2}$ in this case, i.e. the SI cannot be violated.

Finally, we prove that even if the eavesdropper has the ability to control the measurement results of $N-1$ participants, the SI of Eq. (\ref{nsi}) still cannot be violated. We first study the value of $v^{\pm}_{k}=(-1)^{[k(k\pm1)/2]}$. We assume $k=4l+m$, $m$ is the remainder of $k$ divided by $4$, so
\begin{eqnarray}
v^{+}_{k}&=&(-1)^{[k(k+1)/2]}=(-1)^{[(4l+m)(4l+m+1)/2]}\nonumber\\
&=&\left\{
          \begin{aligned}
          1   \quad m=0,3\\
          -1  \quad m=1,2\\
          \end{aligned}
          \right
          .
\label{v1}
\end{eqnarray}
\begin{eqnarray}
v^{-}_{k}&=&(-1)^{[k(k-1)/2]}=(-1)^{[(4l+m)(4l+m-1)/2]}\nonumber\\
&=&\left\{
          \begin{aligned}
          1   \quad m=0,1\\
          -1  \quad m=2,3\\
          \end{aligned}
          \right
          .
\label{v2}
\end{eqnarray}
We now study the relation between the values of $S^{+}_{N-1}$ and $S^{-}_{N-1}$, they are not independent since they consist of the same quantities of $A^{(1)}_{x_1}A^{(2)}_{x_2}\cdot\cdot\cdot A^{(N-1)}_{x_{N-1}}$.
From Eq. (\ref{v1}) we know, if we want $S^{+}_{N-1}$ to takes its maximum value of $2^{N-1}$, we must let the quantities of $A^{(1)}_{x_1}A^{(2)}_{x_2}\cdot\cdot\cdot A^{(N-1)}_{x_{N-1}}$ corresponding to $m=0,3$ take the value $1$, and let the quantities of $A^{(1)}_{x_1}A^{(2)}_{x_2}\cdot\cdot\cdot A^{(N-1)}_{x_{N-1}}$ corresponding to $m=1,2$ take the value $-1$.
We assume that in $S^{\pm}_{N-1}$, the numbers of quantities of $A^{(1)}_{x_1}A^{(2)}_{x_2}\cdot\cdot\cdot A^{(N-1)}_{x_{N-1}}$ corresponding to $m=0,1,2,3$ are $M_{0}$, $M_{1}$, $M_{2}$, $M_{3}$ respectively.
So if $S^{+}_{N-1}$ takes its maximum value of $2^{N-1}$, from Eq. (\ref{v2}) we find the value of $S^{-}_{N-1}$ is
\begin{eqnarray}
S^{-}_{N-1}=M_{0}-M_{1}+M_{2}-M_{3}=(1-1)^{N-1}=0.\nonumber\\
\label{ss}
\end{eqnarray}

If we change one quantity of all $A^{(1)}_{x_1}A^{(2)}_{x_2}\cdot\cdot\cdot A^{(N-1)}_{x_{N-1}}$'s to its opposite number, the value of $S^{+}_{N-1}$ becomes $2^{N-1}-2$ and the value of $S^{-}_{N-1}$ becomes $\pm2$.
We can repeat this process until $S^{+}_{N-1}$ equals $0$ and $S^{-}_{N-1}$ equals $\pm2^{N-1}$.
Given all of that, we finally conclude
\begin{eqnarray}
\left|\langle S^{\pm}_{N}\rangle\right|\leq \left|\langle S^{+}_{N-1}\rangle\right|+\left|\langle S^{-}_{N-1}\rangle\right|=2^{N-1}
\end{eqnarray}
So even if the eavesdropper can control the measurement results of $N-1$ participants, he still unable to create a violation of SI. It also means that the eavesdropper is unable to deceive the legitimate users into believing that the key has been successfully distributed.

\section{Secret key rate}
Different from the usual QKD, device-independent quantum key distribution (DIQKD) protocols aim at
establishing the security of key distribution based on the most fundamental assumptions. In DIQKD, legitimate users can not
only completely ignore the internal working of the quantum measurement apparatuses used in the protocol, but also do not
have to have any information about the particles in their hands. In addition to other fundamental assumptions, the security
proof of a DIQKD protocol can be based entirely on the correctness of quantum theory and observable data.
The aim of DIQKD is to design protocols secure against more powerful eavesdroppers than that in usual QKD,
so from the perspective of DIQKD many usual QKD protocols are not secure. For example, if Alice and Bob actually share
four-dimensional particles and the eavesdropper Eve can tamper their quantum measurement apparatuses so that their actual measurements not correspond
to their expected ones, the BB$84$ \cite{bennett1984quantum} protocol is not secure \cite{Pironio_2009}.

Quantum nonlocality is a necessary condition to the security of DIQKD. The first quantitative relation between secret-key rate and
the violation of CHSH inequality was derived by Ac\'{\i}n \emph{et al} \cite{PhysRevLett.98.230501,Pironio_2009}, this initial result was only considered under the collective attacks condition, and it was later extended to the most powerful coherent attacks condition \cite{PhysRevLett.113.140501}. After that, many secret-key rates were
obtained along with different proposed DIQKD protocols whose securities are guaranteed by violations of different quantum nonlocality inequalities \cite{PhysRevLett.108.100402,PhysRevA.97.022307,PhysRevResearch.2.023251}, and many moderate improvements were achieved by considering actual
noisy processing \cite{Woodhead2021deviceindependent,PhysRevLett.124.230502,Sekatski2021deviceindependent} and imperfect detection efficiency \cite{PhysRevLett.128.110506,PhysRevA.103.052436}.
Since quantum nonlocality is a vulnerable resource and the entanglement of quantum states
is often decayed during transmission, we need here to derive the secret-key rate of our proposed protocol under the condition of using
quantum states with non-maximal entanglement. Next we will calculate the secret-key rate in the three-party scenario, we consider eavesdropper applying
the convex combination (CC) attack \cite{Acin2006,PhysRevLett.97.120405,PhysRevLett.127.050503}, and the non-maximal entanglement state we consider is the Werner state \cite{PhysRevA.40.4277}.

We now give a brief introduction to CC attack. Since we assume the source of particles is under the control of the eavesdropper Eve and
three legitimate users announce their measurement choices for every round, Eve has the ability to mimic the legitimate users' correlation as the following
equation
\begin{eqnarray}
&&P_{ABC}(A_{x},B_{y},C_{z}|\hat{A}_{x},\hat{B}_{y},\hat{C}_{z})\nonumber\\
&=&q_{L}P^{L}_{ABC}(A_{x},B_{y},C_{z}|\hat{A}_{x},\hat{B}_{y},\hat{C}_{z})\nonumber\\
&&+(1-q_{L})P^{NL}_{ABC}(A_{x},B_{y},C_{z}|\hat{A}_{x},\hat{B}_{y},\hat{C}_{z}).
\label{cca}
\end{eqnarray}
Where $q_{L}\in[0,1]$, we call it the local weight. These $\hat{A}_{x}$, $\hat{B}_{y}$, and $\hat{C}_{z}$ denote the measurement choices of Alice, Bob, and Carol respectively,
and $A_{x}$, $B_{y}$, $C_{z}$ are their outcomes. We
call $P_{ABC}(A_{x},B_{y},C_{z}|\hat{A}_{x},\hat{B}_{y},\hat{C}_{z})$ is the legitimate users' observed correlation, $P^{L}_{ABC}(A_{x},B_{y},C_{z}|\hat{A}_{x},\hat{B}_{y},\hat{C}_{z})$
is local correlation (in our case of three-party key distribution we require it to be Svetlichny local, i.e., it cannot result in the
violation of SI), $P^{NL}_{ABC}(A_{x},B_{y},C_{z}|\hat{A}_{x},\hat{B}_{y},\hat{C}_{z})$ is a judiciously chosen nonlocal quantum correlation. From Eq. (\ref{cca}) we see that in order to
mimic legitimate users' correlation Eve distributes local deterministic correlations with probability $q_{L}$, and she distributes nonlocal quantum correlation with
probability $1-q_{L}$. Since in those cases of distribution of local deterministic correlations Eve can have complete information of outcomes of legitimate users' measurements,
she certainly wants to maximize $q_{L}$. This is CC attack.

We assume that the targeted state $\ket{\Psi}_{GHZ}$ is affected by white noise during transmission, and it is transformed into the three-qubit Werner state
\begin{eqnarray}
\rho^{v}_{ABC}=v\ket{\Psi}_{GHZ}\bra{\Psi}+(1-v)\frac{\mathbb{I}}{8},
\label{werner}
\end{eqnarray}
we call $v\in[0,1]$ the visibility, only when $v>\frac{1}{\sqrt{2}}$ the measurement protocol Eq.(\ref{protocal3}) can result in the violation of SI, we denote $v_{L}=\frac{1}{\sqrt{2}}$. We denote the correlation which is given rise to by the Werner state $\rho^{v}_{ABC}$ as $P^{v}_{ABC}$. So if the Werner state
 $\rho^{v}_{ABC}$ give rise to the legitimate users' observed correlation $P_{ABC}(A_{x},B_{y},C_{z}|\hat{A}_{x},\hat{B}_{y},\hat{C}_{z})$
in Eq.(\ref{cca}), it is intuitive that in order to maximize $q_{L}$, Eve should use nonlocal correlation $P^{NL}_{ABC}(A_{x},B_{y},C_{z}|\hat{A}_{x},\hat{B}_{y},\hat{C}_{z})=P^{v=1}_{ABC}$
and local correlation $P^{L}_{ABC}(A_{x},B_{y},C_{z}|\hat{A}_{x},\hat{B}_{y},\hat{C}_{z})=P^{v_{L}}_{ABC}$.
It's easy to find the maximum $q_{L}$,
\begin{eqnarray}
\rho^{v}_{ABC}&=&q_{L}\big[v_{L}\ket{\Psi}_{GHZ}\bra{\Psi}+(1-v_{L})\frac{\mathbb{I}}{8}\big]\nonumber\\
&&+(1-q_{L})\ket{\Psi}_{GHZ}\bra{\Psi},
\label{werner2}
\end{eqnarray}
we can obtain $q_{L}=\frac{1-v}{1-v_{L}}$ by comparing Eq.(\ref{werner}) and Eq.(\ref{werner2}).

\begin{figure}[t]
\includegraphics[width=0.95\columnwidth,
height=0.65\columnwidth]{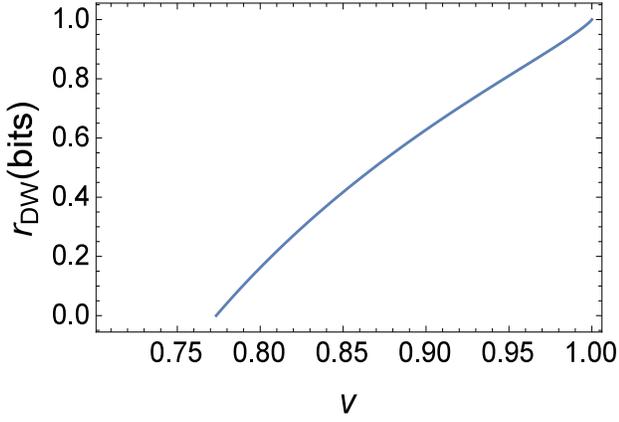} \caption{Extractable secret-key rate $r_{DW}$ against CC attack in the device-independent scenario for the Werner state.
Here, the Werner state is given by Eq.(\ref{werner}) and the $r_{DW}$ is given by Eq.(\ref{key rate 4}). When $v>\frac{1}{2-v_{L}}\approx 0.773459$ the $r_{DW}$ is greater than zero.
The value of $r_{DW}$ monotonously approaches $1$ as the value of visibility $v$ increases, and $r_{DW}=1$ when $v=1$.}
\label{fig3}
\end{figure}

Finally we calculate the secret-key rate of our protocol for the Werner state. In our three-party key distribution protocol, the raw keys are extracted from the following measurement settings
\begin{eqnarray}
&&\big(\hat{A}_{2},\hat{B}_{0},\hat{C}_{0}\big), ~~~~~~~~~~~\big(\hat{A}_{2},\hat{B}_{1},\hat{C}_{1}\big)\nonumber\\
&&\big(\hat{A}_{3},\hat{B}_{0},\hat{C}_{1}\big), ~~~~~~~~~~~\big(\hat{A}_{3},\hat{B}_{1},\hat{C}_{0}\big).
\label{raw key}
\end{eqnarray}
We assume that the probability that the legitimate users choose to make each measurement is equal, and it's obvious that
the secret-key rate calculated from each measurement setting is the same, so we will only calculate the secret-key rate of
the first measurement setting $\big(\hat{A}_{2},\hat{B}_{0},\hat{C}_{0}\big)$.
We use the Devetak-Winter formula \cite{Devetak2005} to calculate the lower bound of the secret-key rate
\begin{eqnarray}
r_{DW}\geq H(A_{2}|E)-H(A_{2}|B_{0},C_{0}).
\label{key rate}
\end{eqnarray}
Where $E$ denote measurement result of Eve, $H(A_{2}|E)$ and $H(A_{2}|B_{0},C_{0})$ both are the conditional Shannon entropy, $H(A_{2}|E)$ quantifies the correlation between Alice and Eve, and
$H(A_{2}|B_{0},C_{0})$ quantifies the correlation between Alice and Bob and Carol. Since three-party key strings have collective correlations rather than pairwise correlations in the two-party case, Eq.(\ref{key rate}) is slightly different from its standard form for the two-party case. If the three legitimate users actually share the
Werner state Eq.(\ref{werner}) which will give rise to correlation $P^{v}_{ABC}$, from Eq.(\ref{werner2}) we can get $P^{v}_{ABC}=q_{L} P^{v_{L}}_{ABC}+(1-q_{L})P^{v=1}_{ABC}$.
Since $P^{v=1}_{ABC}$ is the correlation generated by the maximum entanglement quantum state $\ket{\Psi}_{GHZ}$, the perfect monogamy force that Eve has no information about $A_{2}$ in this case.
We assume that Eve has the power to get complete information about Alice's outcome $A_{2}$ in the case of distribution of local correlation $P^{v_{L}}_{ABC}$. So for every measurement result
$E$ of Eve, we have conditional probability $P(A_{2}=E|E)=\frac{1+q_{L}}{2}$ ($P(A_{2}\neq E|E)=\frac{1-q_{L}}{2}$), and
\begin{eqnarray}
H(A_{2}|E)&=&-\sum_{A_{2},E}{P(E) P(A_{2}|E)\log_{2}{P(A_{2}|E)}}\nonumber\\
&=&h\left(\frac{1+q_{L}}{2}\right),
\label{key rate 2}
\end{eqnarray}
where $h$ is the binary entropy. Then we calculate $H(A_{2}|B_{0},C_{0})$. From the fourth equation of Eq.(\ref{keycalculate}) and Eq.(\ref{werner}), we can obtain
$P(A_{2}=B_{0}C_{0}|B_{0},C_{0})=\frac{1+v}{2}$ ($P(A_{2}\neq B_{0}C_{0}|B_{0},C_{0})=\frac{1-v}{2}$) for every outcomes $B_{0}$ and $C_{0}$, and then
\begin{eqnarray}
&&H(A_{2}|B_{0},C_{0})\nonumber\\
&=&-\sum_{A_{2},B_{0},C_{0}}{P(B_{0},C_{0})P(A_{2}|B_{0},C_{0})\log_{2}{P(A_{2}|B_{0},C_{0})}}\nonumber\\
&=&h\left(\frac{1+v}{2}\right).
\label{key rate 3}
\end{eqnarray}
Substituting these results into  Eq.(\ref{key rate}), we obtain
\begin{eqnarray}
r_{DW}\geq h\left(\frac{1+q_{L}}{2}\right)-h\left(\frac{1+v}{2}\right).
\label{key rate 4}
\end{eqnarray}
Only when $v>\frac{1}{2-v_{L}}\approx 0.773459$ the lower bound of the secret-key rate is greater than zero.
We depict the extractable secret-key rate $r_{DW}$ in
 Fig.\ref{fig3}, and it manifests that as the value of visibility $v$ increases the value of $r_{DW}$ monotonously approaches $1$ and $r_{DW}=1$ when $v=1$.

We notice that the threshold visibility $v=\frac{1}{2-v_{L}}$ for the positive secret-key rate of DIQKD protocols is greater
than the threshold visibility $v_{L}=\frac{1}{\sqrt{2}}$ for the violation of SI, and we have two comments on this result.
First, in our protocol we assume that Eve has the power to get complete information about Alice's outcomes in the case of distribution of local correlation $P^{v_{L}}_{ABC}$, so if, in fact Eve does not have such a powerful ability we will get a less threshold visibility for the positive
secret-key rate, which should be closer to $v_{L}$. Second, even if there exists such a powerful Eve, we can still get a secure DIQKD protocol
as long as $v>\frac{1}{2-v_{L}}$.

\section{Conclusion}
Multipartite private keys have collective correlations rather than pairwise correlations, so no participant can infer the keys of others from the keys in his own hands. Because of this property, there are a lot of tasks in which multipartite keys are applicable but two-partite keys are not competent. In this paper, we present a QKD scheme in which participants can securely share a set of multipartite keys by using a sequence of multi-particle GHZ states. We prove that the violation of SI can be utilized to test for eavesdropping, and even when the eavesdropper can completely control the outcomes of many participants' measurements, our scheme still ensures the device-independent security of the keys distribution.
In the ideal case, the perfect monogamy of the GHZ state guarantee the device-independent security of our protocol,
however quantum entanglement is a vulnerable resource which is often decayed during transmission. So we need to derive the secret-key rate of our protocol under the condition of using quantum states with non-maximal entanglement.
We then calculate the extractable secret-key rate of our three-party key distribution protocol for the Werner state in the device-independent scenario. We find that the value of the extractable secret-key rate monotonously approaches $1$ as the value of the visibility of the Werner state increases, and it reaches its maximum value $1$ when the Werner state becomes the GHZ state.

In the two-party case, if each party has only two inputs and two outputs, there is only one non-trivial quantum nonlocality inequality (CHSH inequality) which can be used to guarantee the security of two-party QKD. Unlike that in the two-party case, there exist many kinds of quantum multipartite correlations. For example, in the case of three-party, there exist $53856$ extremal no-signaling correlations which belong to $46$ inequivalent classes in the case of that each party has two inputs and two outputs \cite{Pironio2011}. So for the multi-party QKD, there exist many different quantum nonlocality inequalities that all can be used to guarantee the security. There are already some multi-party QKD protocols whose security are guaranteed by violations of different quantum nonlocality inequalities \cite{PhysRevLett.108.100402,PhysRevA.97.022307,PhysRevResearch.2.023251}.  In my view, since violations of different quantum nonlocality inequalities display different kinds of quantum multipartite correlations, each inequality has its own advantages as the security guarantee of QKD.

\section*{Acknowledgments}
~The author is grateful to the anonymous referees for their
valuable comments and suggestions to improve the quality
of the paper. This work is supported by the National Natural Science Foundation of China under Grant No. 11005031.

\section*{Data Availability Statement}
~This manuscript has no associated data or the data will not be deposited. [Author’s
comments: All relevant data are in the paper itself.]




\bibliographystyle{apsrev4-1}
\bibliography{multipartite}

\begin{thebibliography}{62}%
\makeatletter
\providecommand \@ifxundefined [1]{%
 \@ifx{#1\undefined}
}%
\providecommand \@ifnum [1]{%
 \ifnum #1\expandafter \@firstoftwo
 \else \expandafter \@secondoftwo
 \fi
}%
\providecommand \@ifx [1]{%
 \ifx #1\expandafter \@firstoftwo
 \else \expandafter \@secondoftwo
 \fi
}%
\providecommand \natexlab [1]{#1}%
\providecommand \enquote  [1]{``#1''}%
\providecommand \bibnamefont  [1]{#1}%
\providecommand \bibfnamefont [1]{#1}%
\providecommand \citenamefont [1]{#1}%
\providecommand \href@noop [0]{\@secondoftwo}%
\providecommand \href [0]{\begingroup \@sanitize@url \@href}%
\providecommand \@href[1]{\@@startlink{#1}\@@href}%
\providecommand \@@href[1]{\endgroup#1\@@endlink}%
\providecommand \@sanitize@url [0]{\catcode `\\12\catcode `\$12\catcode
  `\&12\catcode `\#12\catcode `\^12\catcode `\_12\catcode `\%12\relax}%
\providecommand \@@startlink[1]{}%
\providecommand \@@endlink[0]{}%
\providecommand \url  [0]{\begingroup\@sanitize@url \@url }%
\providecommand \@url [1]{\endgroup\@href {#1}{\urlprefix }}%
\providecommand \urlprefix  [0]{URL }%
\providecommand \Eprint [0]{\href }%
\providecommand \doibase [0]{http://dx.doi.org/}%
\providecommand \selectlanguage [0]{\@gobble}%
\providecommand \bibinfo  [0]{\@secondoftwo}%
\providecommand \bibfield  [0]{\@secondoftwo}%
\providecommand \translation [1]{[#1]}%
\providecommand \BibitemOpen [0]{}%
\providecommand \bibitemStop [0]{}%
\providecommand \bibitemNoStop [0]{.\EOS\space}%
\providecommand \EOS [0]{\spacefactor3000\relax}%
\providecommand \BibitemShut  [1]{\csname bibitem#1\endcsname}%
\let\auto@bib@innerbib\@empty
\bibitem [{\citenamefont {Bennett}\ and\ \citenamefont
  {Brassard}(1984)}]{bennett1984quantum}%
  \BibitemOpen
  \bibfield  {author} {\bibinfo {author} {\bibfnamefont {C.~H.}\ \bibnamefont
  {Bennett}}\ and\ \bibinfo {author} {\bibfnamefont {G.}~\bibnamefont
  {Brassard}},\ }in\ \href@noop {} {\emph {\bibinfo {booktitle} {Proc. IEEE
  Int. Conf. on Computers, Systems and Signal Processing, Bangalore, India}}}\
  (\bibinfo {year} {1984})\ pp.\ \bibinfo {pages} {175--179}\BibitemShut
  {NoStop}%
\bibitem [{\citenamefont {Ekert}(1991)}]{PhysRevLett.67.661}%
  \BibitemOpen
  \bibfield  {author} {\bibinfo {author} {\bibfnamefont {A.~K.}\ \bibnamefont
  {Ekert}},\ }\href {\doibase 10.1103/PhysRevLett.67.661} {\bibfield  {journal}
  {\bibinfo  {journal} {Phys. Rev. Lett.}\ }\textbf {\bibinfo {volume} {67}},\
  \bibinfo {pages} {661} (\bibinfo {year} {1991})}\BibitemShut {NoStop}%
\bibitem [{\citenamefont {Bell}(1964)}]{bell1964einstein}%
  \BibitemOpen
  \bibfield  {author} {\bibinfo {author} {\bibfnamefont {J.~S.}\ \bibnamefont
  {Bell}},\ }\href@noop {} {\bibfield  {journal} {\bibinfo  {journal} {Physics
  Physique Fizika}\ }\textbf {\bibinfo {volume} {1}},\ \bibinfo {pages} {195}
  (\bibinfo {year} {1964})}\BibitemShut {NoStop}%
\bibitem [{\citenamefont {Bell}\ and\ \citenamefont
  {Mermin}(1988)}]{bell1988speakable}%
  \BibitemOpen
  \bibfield  {author} {\bibinfo {author} {\bibfnamefont {J.}~\bibnamefont
  {Bell}}\ and\ \bibinfo {author} {\bibfnamefont {N.~D.}\ \bibnamefont
  {Mermin}},\ }\href@noop {} {\bibfield  {journal} {\bibinfo  {journal}
  {Physics Today}\ }\textbf {\bibinfo {volume} {41}},\ \bibinfo {pages} {89}
  (\bibinfo {year} {1988})}\BibitemShut {NoStop}%
\bibitem [{\citenamefont {Clauser}\ \emph {et~al.}(1969)\citenamefont
  {Clauser}, \citenamefont {Horne}, \citenamefont {Shimony},\ and\
  \citenamefont {Holt}}]{PhysRevLett.23.880}%
  \BibitemOpen
  \bibfield  {author} {\bibinfo {author} {\bibfnamefont {J.~F.}\ \bibnamefont
  {Clauser}}, \bibinfo {author} {\bibfnamefont {M.~A.}\ \bibnamefont {Horne}},
  \bibinfo {author} {\bibfnamefont {A.}~\bibnamefont {Shimony}}, \ and\
  \bibinfo {author} {\bibfnamefont {R.~A.}\ \bibnamefont {Holt}},\ }\href
  {\doibase 10.1103/PhysRevLett.23.880} {\bibfield  {journal} {\bibinfo
  {journal} {Phys. Rev. Lett.}\ }\textbf {\bibinfo {volume} {23}},\ \bibinfo
  {pages} {880} (\bibinfo {year} {1969})}\BibitemShut {NoStop}%
\bibitem [{\citenamefont {Bennett}\ \emph
  {et~al.}(1992{\natexlab{a}})\citenamefont {Bennett}, \citenamefont
  {Brassard},\ and\ \citenamefont {Mermin}}]{bennett1992quantum}%
  \BibitemOpen
  \bibfield  {author} {\bibinfo {author} {\bibfnamefont {C.~H.}\ \bibnamefont
  {Bennett}}, \bibinfo {author} {\bibfnamefont {G.}~\bibnamefont {Brassard}}, \
  and\ \bibinfo {author} {\bibfnamefont {N.~D.}\ \bibnamefont {Mermin}},\
  }\href@noop {} {\bibfield  {journal} {\bibinfo  {journal} {Physical review
  letters}\ }\textbf {\bibinfo {volume} {68}},\ \bibinfo {pages} {557}
  (\bibinfo {year} {1992}{\natexlab{a}})}\BibitemShut {NoStop}%
\bibitem [{\citenamefont {Einstein}\ \emph {et~al.}(1935)\citenamefont
  {Einstein}, \citenamefont {Podolsky},\ and\ \citenamefont
  {Rosen}}]{PhysRev.47.777}%
  \BibitemOpen
  \bibfield  {author} {\bibinfo {author} {\bibfnamefont {A.}~\bibnamefont
  {Einstein}}, \bibinfo {author} {\bibfnamefont {B.}~\bibnamefont {Podolsky}},
  \ and\ \bibinfo {author} {\bibfnamefont {N.}~\bibnamefont {Rosen}},\ }\href
  {\doibase 10.1103/PhysRev.47.777} {\bibfield  {journal} {\bibinfo  {journal}
  {Phys. Rev.}\ }\textbf {\bibinfo {volume} {47}},\ \bibinfo {pages} {777}
  (\bibinfo {year} {1935})}\BibitemShut {NoStop}%
\bibitem [{\citenamefont {Bennett}\ \emph
  {et~al.}(1992{\natexlab{b}})\citenamefont {Bennett}, \citenamefont
  {Brassard},\ and\ \citenamefont {Mermin}}]{PhysRevLett.68.557}%
  \BibitemOpen
  \bibfield  {author} {\bibinfo {author} {\bibfnamefont {C.~H.}\ \bibnamefont
  {Bennett}}, \bibinfo {author} {\bibfnamefont {G.}~\bibnamefont {Brassard}}, \
  and\ \bibinfo {author} {\bibfnamefont {N.~D.}\ \bibnamefont {Mermin}},\
  }\href {\doibase 10.1103/PhysRevLett.68.557} {\bibfield  {journal} {\bibinfo
  {journal} {Phys. Rev. Lett.}\ }\textbf {\bibinfo {volume} {68}},\ \bibinfo
  {pages} {557} (\bibinfo {year} {1992}{\natexlab{b}})}\BibitemShut {NoStop}%
\bibitem [{\citenamefont {Gisin}\ \emph {et~al.}(2002)\citenamefont {Gisin},
  \citenamefont {Ribordy}, \citenamefont {Tittel},\ and\ \citenamefont
  {Zbinden}}]{RevModPhys.74.145}%
  \BibitemOpen
  \bibfield  {author} {\bibinfo {author} {\bibfnamefont {N.}~\bibnamefont
  {Gisin}}, \bibinfo {author} {\bibfnamefont {G.}~\bibnamefont {Ribordy}},
  \bibinfo {author} {\bibfnamefont {W.}~\bibnamefont {Tittel}}, \ and\ \bibinfo
  {author} {\bibfnamefont {H.}~\bibnamefont {Zbinden}},\ }\href {\doibase
  10.1103/RevModPhys.74.145} {\bibfield  {journal} {\bibinfo  {journal} {Rev.
  Mod. Phys.}\ }\textbf {\bibinfo {volume} {74}},\ \bibinfo {pages} {145}
  (\bibinfo {year} {2002})}\BibitemShut {NoStop}%
\bibitem [{\citenamefont {Gisin}\ and\ \citenamefont
  {Huttner}(1997)}]{gisin1997quantum}%
  \BibitemOpen
  \bibfield  {author} {\bibinfo {author} {\bibfnamefont {N.}~\bibnamefont
  {Gisin}}\ and\ \bibinfo {author} {\bibfnamefont {B.}~\bibnamefont
  {Huttner}},\ }\href@noop {} {\bibfield  {journal} {\bibinfo  {journal}
  {Physics Letters A}\ }\textbf {\bibinfo {volume} {228}},\ \bibinfo {pages}
  {13} (\bibinfo {year} {1997})}\BibitemShut {NoStop}%
\bibitem [{\citenamefont {Scarani}\ and\ \citenamefont
  {Gisin}(2001{\natexlab{a}})}]{PhysRevLett.87.117901}%
  \BibitemOpen
  \bibfield  {author} {\bibinfo {author} {\bibfnamefont {V.}~\bibnamefont
  {Scarani}}\ and\ \bibinfo {author} {\bibfnamefont {N.}~\bibnamefont
  {Gisin}},\ }\href {\doibase 10.1103/PhysRevLett.87.117901} {\bibfield
  {journal} {\bibinfo  {journal} {Phys. Rev. Lett.}\ }\textbf {\bibinfo
  {volume} {87}},\ \bibinfo {pages} {117901} (\bibinfo {year}
  {2001}{\natexlab{a}})}\BibitemShut {NoStop}%
\bibitem [{\citenamefont {Scarani}\ and\ \citenamefont
  {Gisin}(2001{\natexlab{b}})}]{PhysRevA.65.012311}%
  \BibitemOpen
  \bibfield  {author} {\bibinfo {author} {\bibfnamefont {V.}~\bibnamefont
  {Scarani}}\ and\ \bibinfo {author} {\bibfnamefont {N.}~\bibnamefont
  {Gisin}},\ }\href {\doibase 10.1103/PhysRevA.65.012311} {\bibfield  {journal}
  {\bibinfo  {journal} {Phys. Rev. A}\ }\textbf {\bibinfo {volume} {65}},\
  \bibinfo {pages} {012311} (\bibinfo {year} {2001}{\natexlab{b}})}\BibitemShut
  {NoStop}%
\bibitem [{\citenamefont {Cabello}(2001{\natexlab{a}})}]{PhysRevLett.86.1911}%
  \BibitemOpen
  \bibfield  {author} {\bibinfo {author} {\bibfnamefont {A.}~\bibnamefont
  {Cabello}},\ }\href {\doibase 10.1103/PhysRevLett.86.1911} {\bibfield
  {journal} {\bibinfo  {journal} {Phys. Rev. Lett.}\ }\textbf {\bibinfo
  {volume} {86}},\ \bibinfo {pages} {1911} (\bibinfo {year}
  {2001}{\natexlab{a}})}\BibitemShut {NoStop}%
\bibitem [{\citenamefont
  {Cabello}(2001{\natexlab{b}})}]{PhysRevLett.87.010403}%
  \BibitemOpen
  \bibfield  {author} {\bibinfo {author} {\bibfnamefont {A.}~\bibnamefont
  {Cabello}},\ }\href {\doibase 10.1103/PhysRevLett.87.010403} {\bibfield
  {journal} {\bibinfo  {journal} {Phys. Rev. Lett.}\ }\textbf {\bibinfo
  {volume} {87}},\ \bibinfo {pages} {010403} (\bibinfo {year}
  {2001}{\natexlab{b}})}\BibitemShut {NoStop}%
\bibitem [{\citenamefont {Chen}\ \emph {et~al.}(2003)\citenamefont {Chen},
  \citenamefont {Pan}, \citenamefont {Zhang}, \citenamefont {Brukner},\ and\
  \citenamefont {Zeilinger}}]{PhysRevLett.90.160408}%
  \BibitemOpen
  \bibfield  {author} {\bibinfo {author} {\bibfnamefont {Z.-B.}\ \bibnamefont
  {Chen}}, \bibinfo {author} {\bibfnamefont {J.-W.}\ \bibnamefont {Pan}},
  \bibinfo {author} {\bibfnamefont {Y.-D.}\ \bibnamefont {Zhang}}, \bibinfo
  {author} {\bibfnamefont {v.}~\bibnamefont {Brukner}}, \ and\ \bibinfo
  {author} {\bibfnamefont {A.}~\bibnamefont {Zeilinger}},\ }\href {\doibase
  10.1103/PhysRevLett.90.160408} {\bibfield  {journal} {\bibinfo  {journal}
  {Phys. Rev. Lett.}\ }\textbf {\bibinfo {volume} {90}},\ \bibinfo {pages}
  {160408} (\bibinfo {year} {2003})}\BibitemShut {NoStop}%
\bibitem [{\citenamefont {Beige}\ \emph {et~al.}(2002)\citenamefont {Beige},
  \citenamefont {Englert}, \citenamefont {Kurtsiefer},\ and\ \citenamefont
  {Weinfurter}}]{beige2002secure}%
  \BibitemOpen
  \bibfield  {author} {\bibinfo {author} {\bibfnamefont {A.}~\bibnamefont
  {Beige}}, \bibinfo {author} {\bibfnamefont {B.-G.}\ \bibnamefont {Englert}},
  \bibinfo {author} {\bibfnamefont {C.}~\bibnamefont {Kurtsiefer}}, \ and\
  \bibinfo {author} {\bibfnamefont {H.}~\bibnamefont {Weinfurter}},\
  }\href@noop {} {\bibfield  {journal} {\bibinfo  {journal} {Journal of Physics
  A: Mathematical and General}\ }\textbf {\bibinfo {volume} {35}},\ \bibinfo
  {pages} {L407} (\bibinfo {year} {2002})}\BibitemShut {NoStop}%
\bibitem [{\citenamefont {Durt}\ \emph {et~al.}(2003)\citenamefont {Durt},
  \citenamefont {Cerf}, \citenamefont {Gisin},\ and\ \citenamefont
  {\ifmmode~\dot{Z}\else \.{Z}\fi{}ukowski}}]{PhysRevA.67.012311}%
  \BibitemOpen
  \bibfield  {author} {\bibinfo {author} {\bibfnamefont {T.}~\bibnamefont
  {Durt}}, \bibinfo {author} {\bibfnamefont {N.~J.}\ \bibnamefont {Cerf}},
  \bibinfo {author} {\bibfnamefont {N.}~\bibnamefont {Gisin}}, \ and\ \bibinfo
  {author} {\bibfnamefont {M.}~\bibnamefont {\ifmmode~\dot{Z}\else
  \.{Z}\fi{}ukowski}},\ }\href {\doibase 10.1103/PhysRevA.67.012311} {\bibfield
   {journal} {\bibinfo  {journal} {Phys. Rev. A}\ }\textbf {\bibinfo {volume}
  {67}},\ \bibinfo {pages} {012311} (\bibinfo {year} {2003})}\BibitemShut
  {NoStop}%
\bibitem [{\citenamefont {Lo}\ \emph {et~al.}(2005)\citenamefont {Lo},
  \citenamefont {Ma},\ and\ \citenamefont {Chen}}]{PhysRevLett.94.230504}%
  \BibitemOpen
  \bibfield  {author} {\bibinfo {author} {\bibfnamefont {H.-K.}\ \bibnamefont
  {Lo}}, \bibinfo {author} {\bibfnamefont {X.}~\bibnamefont {Ma}}, \ and\
  \bibinfo {author} {\bibfnamefont {K.}~\bibnamefont {Chen}},\ }\href {\doibase
  10.1103/PhysRevLett.94.230504} {\bibfield  {journal} {\bibinfo  {journal}
  {Phys. Rev. Lett.}\ }\textbf {\bibinfo {volume} {94}},\ \bibinfo {pages}
  {230504} (\bibinfo {year} {2005})}\BibitemShut {NoStop}%
\bibitem [{\citenamefont {Scarani}\ \emph {et~al.}(2009)\citenamefont
  {Scarani}, \citenamefont {Bechmann-Pasquinucci}, \citenamefont {Cerf},
  \citenamefont {Du\ifmmode~\check{s}\else \v{s}\fi{}ek}, \citenamefont
  {L\"utkenhaus},\ and\ \citenamefont {Peev}}]{RevModPhys.81.1301}%
  \BibitemOpen
  \bibfield  {author} {\bibinfo {author} {\bibfnamefont {V.}~\bibnamefont
  {Scarani}}, \bibinfo {author} {\bibfnamefont {H.}~\bibnamefont
  {Bechmann-Pasquinucci}}, \bibinfo {author} {\bibfnamefont {N.~J.}\
  \bibnamefont {Cerf}}, \bibinfo {author} {\bibfnamefont {M.}~\bibnamefont
  {Du\ifmmode~\check{s}\else \v{s}\fi{}ek}}, \bibinfo {author} {\bibfnamefont
  {N.}~\bibnamefont {L\"utkenhaus}}, \ and\ \bibinfo {author} {\bibfnamefont
  {M.}~\bibnamefont {Peev}},\ }\href {\doibase 10.1103/RevModPhys.81.1301}
  {\bibfield  {journal} {\bibinfo  {journal} {Rev. Mod. Phys.}\ }\textbf
  {\bibinfo {volume} {81}},\ \bibinfo {pages} {1301} (\bibinfo {year}
  {2009})}\BibitemShut {NoStop}%
\bibitem [{\citenamefont {Bera}\ \emph {et~al.}(2016)\citenamefont {Bera},
  \citenamefont {Kumar}, \citenamefont {Rakshit}, \citenamefont {Prabhu},
  \citenamefont {Sen(De)},\ and\ \citenamefont {Sen}}]{PhysRevA.93.032338}%
  \BibitemOpen
  \bibfield  {author} {\bibinfo {author} {\bibfnamefont {A.}~\bibnamefont
  {Bera}}, \bibinfo {author} {\bibfnamefont {A.}~\bibnamefont {Kumar}},
  \bibinfo {author} {\bibfnamefont {D.}~\bibnamefont {Rakshit}}, \bibinfo
  {author} {\bibfnamefont {R.}~\bibnamefont {Prabhu}}, \bibinfo {author}
  {\bibfnamefont {A.}~\bibnamefont {Sen(De)}}, \ and\ \bibinfo {author}
  {\bibfnamefont {U.}~\bibnamefont {Sen}},\ }\href {\doibase
  10.1103/PhysRevA.93.032338} {\bibfield  {journal} {\bibinfo  {journal} {Phys.
  Rev. A}\ }\textbf {\bibinfo {volume} {93}},\ \bibinfo {pages} {032338}
  (\bibinfo {year} {2016})}\BibitemShut {NoStop}%
\bibitem [{\citenamefont {Yang}\ \emph {et~al.}(2016)\citenamefont {Yang},
  \citenamefont {Wei}, \citenamefont {Ma}, \citenamefont {Sun}, \citenamefont
  {Liu}, \citenamefont {Yin}, \citenamefont {Li}, \citenamefont {Lian},
  \citenamefont {Du},\ and\ \citenamefont {Wu}}]{PhysRevA.93.052303}%
  \BibitemOpen
  \bibfield  {author} {\bibinfo {author} {\bibfnamefont {X.}~\bibnamefont
  {Yang}}, \bibinfo {author} {\bibfnamefont {K.}~\bibnamefont {Wei}}, \bibinfo
  {author} {\bibfnamefont {H.}~\bibnamefont {Ma}}, \bibinfo {author}
  {\bibfnamefont {S.}~\bibnamefont {Sun}}, \bibinfo {author} {\bibfnamefont
  {H.}~\bibnamefont {Liu}}, \bibinfo {author} {\bibfnamefont {Z.}~\bibnamefont
  {Yin}}, \bibinfo {author} {\bibfnamefont {Z.}~\bibnamefont {Li}}, \bibinfo
  {author} {\bibfnamefont {S.}~\bibnamefont {Lian}}, \bibinfo {author}
  {\bibfnamefont {Y.}~\bibnamefont {Du}}, \ and\ \bibinfo {author}
  {\bibfnamefont {L.}~\bibnamefont {Wu}},\ }\href {\doibase
  10.1103/PhysRevA.93.052303} {\bibfield  {journal} {\bibinfo  {journal} {Phys.
  Rev. A}\ }\textbf {\bibinfo {volume} {93}},\ \bibinfo {pages} {052303}
  (\bibinfo {year} {2016})}\BibitemShut {NoStop}%
\bibitem [{\citenamefont {Xiang}\ \emph {et~al.}(2017)\citenamefont {Xiang},
  \citenamefont {Kogias}, \citenamefont {Adesso},\ and\ \citenamefont
  {He}}]{PhysRevA.95.010101}%
  \BibitemOpen
  \bibfield  {author} {\bibinfo {author} {\bibfnamefont {Y.}~\bibnamefont
  {Xiang}}, \bibinfo {author} {\bibfnamefont {I.}~\bibnamefont {Kogias}},
  \bibinfo {author} {\bibfnamefont {G.}~\bibnamefont {Adesso}}, \ and\ \bibinfo
  {author} {\bibfnamefont {Q.}~\bibnamefont {He}},\ }\href {\doibase
  10.1103/PhysRevA.95.010101} {\bibfield  {journal} {\bibinfo  {journal} {Phys.
  Rev. A}\ }\textbf {\bibinfo {volume} {95}},\ \bibinfo {pages} {010101}
  (\bibinfo {year} {2017})}\BibitemShut {NoStop}%
\bibitem [{\citenamefont {Zhang}\ \emph {et~al.}(2017)\citenamefont {Zhang},
  \citenamefont {Ding}, \citenamefont {Sheng}, \citenamefont {Zhou},
  \citenamefont {Shi},\ and\ \citenamefont {Guo}}]{PhysRevLett.118.220501}%
  \BibitemOpen
  \bibfield  {author} {\bibinfo {author} {\bibfnamefont {W.}~\bibnamefont
  {Zhang}}, \bibinfo {author} {\bibfnamefont {D.-S.}\ \bibnamefont {Ding}},
  \bibinfo {author} {\bibfnamefont {Y.-B.}\ \bibnamefont {Sheng}}, \bibinfo
  {author} {\bibfnamefont {L.}~\bibnamefont {Zhou}}, \bibinfo {author}
  {\bibfnamefont {B.-S.}\ \bibnamefont {Shi}}, \ and\ \bibinfo {author}
  {\bibfnamefont {G.-C.}\ \bibnamefont {Guo}},\ }\href {\doibase
  10.1103/PhysRevLett.118.220501} {\bibfield  {journal} {\bibinfo  {journal}
  {Phys. Rev. Lett.}\ }\textbf {\bibinfo {volume} {118}},\ \bibinfo {pages}
  {220501} (\bibinfo {year} {2017})}\BibitemShut {NoStop}%
\bibitem [{\citenamefont {Kravtsov}\ and\ \citenamefont
  {Molotkov}(2019)}]{PhysRevA.100.042329}%
  \BibitemOpen
  \bibfield  {author} {\bibinfo {author} {\bibfnamefont {K.~S.}\ \bibnamefont
  {Kravtsov}}\ and\ \bibinfo {author} {\bibfnamefont {S.~N.}\ \bibnamefont
  {Molotkov}},\ }\href {\doibase 10.1103/PhysRevA.100.042329} {\bibfield
  {journal} {\bibinfo  {journal} {Phys. Rev. A}\ }\textbf {\bibinfo {volume}
  {100}},\ \bibinfo {pages} {042329} (\bibinfo {year} {2019})}\BibitemShut
  {NoStop}%
\bibitem [{\citenamefont {Yin}\ \emph {et~al.}(2017)\citenamefont {Yin},
  \citenamefont {Cao}, \citenamefont {Li}, \citenamefont {Ren}, \citenamefont
  {Liao}, \citenamefont {Zhang}, \citenamefont {Cai}, \citenamefont {Liu},
  \citenamefont {Li}, \citenamefont {Dai}, \citenamefont {Li}, \citenamefont
  {Huang}, \citenamefont {Deng}, \citenamefont {Li}, \citenamefont {Zhang},
  \citenamefont {Liu}, \citenamefont {Chen}, \citenamefont {Lu}, \citenamefont
  {Shu}, \citenamefont {Peng}, \citenamefont {Wang},\ and\ \citenamefont
  {Pan}}]{PhysRevLett.119.200501}%
  \BibitemOpen
  \bibfield  {author} {\bibinfo {author} {\bibfnamefont {J.}~\bibnamefont
  {Yin}}, \bibinfo {author} {\bibfnamefont {Y.}~\bibnamefont {Cao}}, \bibinfo
  {author} {\bibfnamefont {Y.-H.}\ \bibnamefont {Li}}, \bibinfo {author}
  {\bibfnamefont {J.-G.}\ \bibnamefont {Ren}}, \bibinfo {author} {\bibfnamefont
  {S.-K.}\ \bibnamefont {Liao}}, \bibinfo {author} {\bibfnamefont
  {L.}~\bibnamefont {Zhang}}, \bibinfo {author} {\bibfnamefont {W.-Q.}\
  \bibnamefont {Cai}}, \bibinfo {author} {\bibfnamefont {W.-Y.}\ \bibnamefont
  {Liu}}, \bibinfo {author} {\bibfnamefont {B.}~\bibnamefont {Li}}, \bibinfo
  {author} {\bibfnamefont {H.}~\bibnamefont {Dai}}, \bibinfo {author}
  {\bibfnamefont {M.}~\bibnamefont {Li}}, \bibinfo {author} {\bibfnamefont
  {Y.-M.}\ \bibnamefont {Huang}}, \bibinfo {author} {\bibfnamefont
  {L.}~\bibnamefont {Deng}}, \bibinfo {author} {\bibfnamefont {L.}~\bibnamefont
  {Li}}, \bibinfo {author} {\bibfnamefont {Q.}~\bibnamefont {Zhang}}, \bibinfo
  {author} {\bibfnamefont {N.-L.}\ \bibnamefont {Liu}}, \bibinfo {author}
  {\bibfnamefont {Y.-A.}\ \bibnamefont {Chen}}, \bibinfo {author}
  {\bibfnamefont {C.-Y.}\ \bibnamefont {Lu}}, \bibinfo {author} {\bibfnamefont
  {R.}~\bibnamefont {Shu}}, \bibinfo {author} {\bibfnamefont {C.-Z.}\
  \bibnamefont {Peng}}, \bibinfo {author} {\bibfnamefont {J.-Y.}\ \bibnamefont
  {Wang}}, \ and\ \bibinfo {author} {\bibfnamefont {J.-W.}\ \bibnamefont
  {Pan}},\ }\href {\doibase 10.1103/PhysRevLett.119.200501} {\bibfield
  {journal} {\bibinfo  {journal} {Phys. Rev. Lett.}\ }\textbf {\bibinfo
  {volume} {119}},\ \bibinfo {pages} {200501} (\bibinfo {year}
  {2017})}\BibitemShut {NoStop}%
\bibitem [{\citenamefont {Pirandola}\ \emph {et~al.}(2020)\citenamefont
  {Pirandola}, \citenamefont {Andersen}, \citenamefont {Banchi}, \citenamefont
  {Berta}, \citenamefont {Bunandar}, \citenamefont {Colbeck}, \citenamefont
  {Englund}, \citenamefont {Gehring}, \citenamefont {Lupo}, \citenamefont
  {Ottaviani} \emph {et~al.}}]{pirandola2020advances}%
  \BibitemOpen
  \bibfield  {author} {\bibinfo {author} {\bibfnamefont {S.}~\bibnamefont
  {Pirandola}}, \bibinfo {author} {\bibfnamefont {U.~L.}\ \bibnamefont
  {Andersen}}, \bibinfo {author} {\bibfnamefont {L.}~\bibnamefont {Banchi}},
  \bibinfo {author} {\bibfnamefont {M.}~\bibnamefont {Berta}}, \bibinfo
  {author} {\bibfnamefont {D.}~\bibnamefont {Bunandar}}, \bibinfo {author}
  {\bibfnamefont {R.}~\bibnamefont {Colbeck}}, \bibinfo {author} {\bibfnamefont
  {D.}~\bibnamefont {Englund}}, \bibinfo {author} {\bibfnamefont
  {T.}~\bibnamefont {Gehring}}, \bibinfo {author} {\bibfnamefont
  {C.}~\bibnamefont {Lupo}}, \bibinfo {author} {\bibfnamefont {C.}~\bibnamefont
  {Ottaviani}},  \emph {et~al.},\ }\href@noop {} {\bibfield  {journal}
  {\bibinfo  {journal} {Advances in optics and photonics}\ }\textbf {\bibinfo
  {volume} {12}},\ \bibinfo {pages} {1012} (\bibinfo {year}
  {2020})}\BibitemShut {NoStop}%
\bibitem [{\citenamefont {Brito}\ \emph {et~al.}(2021)\citenamefont {Brito},
  \citenamefont {Canabarro}, \citenamefont {Cavalcanti},\ and\ \citenamefont
  {Chaves}}]{PRXQuantum.2.010304}%
  \BibitemOpen
  \bibfield  {author} {\bibinfo {author} {\bibfnamefont {S.}~\bibnamefont
  {Brito}}, \bibinfo {author} {\bibfnamefont {A.}~\bibnamefont {Canabarro}},
  \bibinfo {author} {\bibfnamefont {D.}~\bibnamefont {Cavalcanti}}, \ and\
  \bibinfo {author} {\bibfnamefont {R.}~\bibnamefont {Chaves}},\ }\href
  {\doibase 10.1103/PRXQuantum.2.010304} {\bibfield  {journal} {\bibinfo
  {journal} {PRX Quantum}\ }\textbf {\bibinfo {volume} {2}},\ \bibinfo {pages}
  {010304} (\bibinfo {year} {2021})}\BibitemShut {NoStop}%
\bibitem [{\citenamefont {Fitzke}\ \emph {et~al.}(2022)\citenamefont {Fitzke},
  \citenamefont {Bialowons}, \citenamefont {Dolejsky}, \citenamefont
  {Tippmann}, \citenamefont {Nikiforov}, \citenamefont {Walther}, \citenamefont
  {Wissel},\ and\ \citenamefont {Gunkel}}]{PRXQuantum.3.020341}%
  \BibitemOpen
  \bibfield  {author} {\bibinfo {author} {\bibfnamefont {E.}~\bibnamefont
  {Fitzke}}, \bibinfo {author} {\bibfnamefont {L.}~\bibnamefont {Bialowons}},
  \bibinfo {author} {\bibfnamefont {T.}~\bibnamefont {Dolejsky}}, \bibinfo
  {author} {\bibfnamefont {M.}~\bibnamefont {Tippmann}}, \bibinfo {author}
  {\bibfnamefont {O.}~\bibnamefont {Nikiforov}}, \bibinfo {author}
  {\bibfnamefont {T.}~\bibnamefont {Walther}}, \bibinfo {author} {\bibfnamefont
  {F.}~\bibnamefont {Wissel}}, \ and\ \bibinfo {author} {\bibfnamefont
  {M.}~\bibnamefont {Gunkel}},\ }\href {\doibase 10.1103/PRXQuantum.3.020341}
  {\bibfield  {journal} {\bibinfo  {journal} {PRX Quantum}\ }\textbf {\bibinfo
  {volume} {3}},\ \bibinfo {pages} {020341} (\bibinfo {year}
  {2022})}\BibitemShut {NoStop}%
\bibitem [{\citenamefont {Bancal}\ \emph
  {et~al.}(2011{\natexlab{a}})\citenamefont {Bancal}, \citenamefont {Brunner},
  \citenamefont {Gisin},\ and\ \citenamefont {Liang}}]{PhysRevLett.106.020405}%
  \BibitemOpen
  \bibfield  {author} {\bibinfo {author} {\bibfnamefont {J.-D.}\ \bibnamefont
  {Bancal}}, \bibinfo {author} {\bibfnamefont {N.}~\bibnamefont {Brunner}},
  \bibinfo {author} {\bibfnamefont {N.}~\bibnamefont {Gisin}}, \ and\ \bibinfo
  {author} {\bibfnamefont {Y.-C.}\ \bibnamefont {Liang}},\ }\href {\doibase
  10.1103/PhysRevLett.106.020405} {\bibfield  {journal} {\bibinfo  {journal}
  {Phys. Rev. Lett.}\ }\textbf {\bibinfo {volume} {106}},\ \bibinfo {pages}
  {020405} (\bibinfo {year} {2011}{\natexlab{a}})}\BibitemShut {NoStop}%
\bibitem [{\citenamefont {Bancal}\ \emph
  {et~al.}(2011{\natexlab{b}})\citenamefont {Bancal}, \citenamefont {Gisin},
  \citenamefont {Liang},\ and\ \citenamefont
  {Pironio}}]{PhysRevLett.106.250404}%
  \BibitemOpen
  \bibfield  {author} {\bibinfo {author} {\bibfnamefont {J.-D.}\ \bibnamefont
  {Bancal}}, \bibinfo {author} {\bibfnamefont {N.}~\bibnamefont {Gisin}},
  \bibinfo {author} {\bibfnamefont {Y.-C.}\ \bibnamefont {Liang}}, \ and\
  \bibinfo {author} {\bibfnamefont {S.}~\bibnamefont {Pironio}},\ }\href
  {\doibase 10.1103/PhysRevLett.106.250404} {\bibfield  {journal} {\bibinfo
  {journal} {Phys. Rev. Lett.}\ }\textbf {\bibinfo {volume} {106}},\ \bibinfo
  {pages} {250404} (\bibinfo {year} {2011}{\natexlab{b}})}\BibitemShut
  {NoStop}%
\bibitem [{\citenamefont {\ifmmode~\dot{Z}\else \.{Z}\fi{}ukowski}\ and\
  \citenamefont {Brukner}(2002)}]{PhysRevLett.88.210401}%
  \BibitemOpen
  \bibfield  {author} {\bibinfo {author} {\bibfnamefont {M.}~\bibnamefont
  {\ifmmode~\dot{Z}\else \.{Z}\fi{}ukowski}}\ and\ \bibinfo {author}
  {\bibfnamefont {C.}~\bibnamefont {Brukner}},\ }\href {\doibase
  10.1103/PhysRevLett.88.210401} {\bibfield  {journal} {\bibinfo  {journal}
  {Phys. Rev. Lett.}\ }\textbf {\bibinfo {volume} {88}},\ \bibinfo {pages}
  {210401} (\bibinfo {year} {2002})}\BibitemShut {NoStop}%
\bibitem [{\citenamefont {Gallego}\ \emph {et~al.}(2011)\citenamefont
  {Gallego}, \citenamefont {W\"urflinger}, \citenamefont {Ac\'{\i}n},\ and\
  \citenamefont {Navascu\'es}}]{PhysRevLett.107.210403}%
  \BibitemOpen
  \bibfield  {author} {\bibinfo {author} {\bibfnamefont {R.}~\bibnamefont
  {Gallego}}, \bibinfo {author} {\bibfnamefont {L.~E.}\ \bibnamefont
  {W\"urflinger}}, \bibinfo {author} {\bibfnamefont {A.}~\bibnamefont
  {Ac\'{\i}n}}, \ and\ \bibinfo {author} {\bibfnamefont {M.}~\bibnamefont
  {Navascu\'es}},\ }\href {\doibase 10.1103/PhysRevLett.107.210403} {\bibfield
  {journal} {\bibinfo  {journal} {Phys. Rev. Lett.}\ }\textbf {\bibinfo
  {volume} {107}},\ \bibinfo {pages} {210403} (\bibinfo {year}
  {2011})}\BibitemShut {NoStop}%
\bibitem [{\citenamefont {Sainz}\ \emph {et~al.}(2014)\citenamefont {Sainz},
  \citenamefont {Fritz}, \citenamefont {Augusiak}, \citenamefont {Brask},
  \citenamefont {Chaves}, \citenamefont {Leverrier},\ and\ \citenamefont
  {Ac\'{\i}n}}]{PhysRevA.89.032117}%
  \BibitemOpen
  \bibfield  {author} {\bibinfo {author} {\bibfnamefont {A.~B.}\ \bibnamefont
  {Sainz}}, \bibinfo {author} {\bibfnamefont {T.}~\bibnamefont {Fritz}},
  \bibinfo {author} {\bibfnamefont {R.}~\bibnamefont {Augusiak}}, \bibinfo
  {author} {\bibfnamefont {J.~B.}\ \bibnamefont {Brask}}, \bibinfo {author}
  {\bibfnamefont {R.}~\bibnamefont {Chaves}}, \bibinfo {author} {\bibfnamefont
  {A.}~\bibnamefont {Leverrier}}, \ and\ \bibinfo {author} {\bibfnamefont
  {A.}~\bibnamefont {Ac\'{\i}n}},\ }\href {\doibase 10.1103/PhysRevA.89.032117}
  {\bibfield  {journal} {\bibinfo  {journal} {Phys. Rev. A}\ }\textbf {\bibinfo
  {volume} {89}},\ \bibinfo {pages} {032117} (\bibinfo {year}
  {2014})}\BibitemShut {NoStop}%
\bibitem [{\citenamefont {Brunner}\ \emph {et~al.}(2014)\citenamefont
  {Brunner}, \citenamefont {Cavalcanti}, \citenamefont {Pironio}, \citenamefont
  {Scarani},\ and\ \citenamefont {Wehner}}]{RevModPhys.86.419}%
  \BibitemOpen
  \bibfield  {author} {\bibinfo {author} {\bibfnamefont {N.}~\bibnamefont
  {Brunner}}, \bibinfo {author} {\bibfnamefont {D.}~\bibnamefont {Cavalcanti}},
  \bibinfo {author} {\bibfnamefont {S.}~\bibnamefont {Pironio}}, \bibinfo
  {author} {\bibfnamefont {V.}~\bibnamefont {Scarani}}, \ and\ \bibinfo
  {author} {\bibfnamefont {S.}~\bibnamefont {Wehner}},\ }\href {\doibase
  10.1103/RevModPhys.86.419} {\bibfield  {journal} {\bibinfo  {journal} {Rev.
  Mod. Phys.}\ }\textbf {\bibinfo {volume} {86}},\ \bibinfo {pages} {419}
  (\bibinfo {year} {2014})}\BibitemShut {NoStop}%
\bibitem [{\citenamefont {Xiang}\ and\ \citenamefont
  {Ren}(2011)}]{xiang2011bound}%
  \BibitemOpen
  \bibfield  {author} {\bibinfo {author} {\bibfnamefont {Y.}~\bibnamefont
  {Xiang}}\ and\ \bibinfo {author} {\bibfnamefont {W.}~\bibnamefont {Ren}},\
  }\href {\doibase https://doi.org/10.26421/QIC11.11-12-5} {\bibfield
  {journal} {\bibinfo  {journal} {Quantum information and computation}\
  }\textbf {\bibinfo {volume} {11}},\ \bibinfo {pages} {0948} (\bibinfo {year}
  {2011})}\BibitemShut {NoStop}%
\bibitem [{\citenamefont {Barrett}\ \emph {et~al.}(2005)\citenamefont
  {Barrett}, \citenamefont {Linden}, \citenamefont {Massar}, \citenamefont
  {Pironio}, \citenamefont {Popescu},\ and\ \citenamefont
  {Roberts}}]{PhysRevA.71.022101}%
  \BibitemOpen
  \bibfield  {author} {\bibinfo {author} {\bibfnamefont {J.}~\bibnamefont
  {Barrett}}, \bibinfo {author} {\bibfnamefont {N.}~\bibnamefont {Linden}},
  \bibinfo {author} {\bibfnamefont {S.}~\bibnamefont {Massar}}, \bibinfo
  {author} {\bibfnamefont {S.}~\bibnamefont {Pironio}}, \bibinfo {author}
  {\bibfnamefont {S.}~\bibnamefont {Popescu}}, \ and\ \bibinfo {author}
  {\bibfnamefont {D.}~\bibnamefont {Roberts}},\ }\href {\doibase
  10.1103/PhysRevA.71.022101} {\bibfield  {journal} {\bibinfo  {journal} {Phys.
  Rev. A}\ }\textbf {\bibinfo {volume} {71}},\ \bibinfo {pages} {022101}
  (\bibinfo {year} {2005})}\BibitemShut {NoStop}%
\bibitem [{\citenamefont {Pironio}\ \emph
  {et~al.}(2011{\natexlab{a}})\citenamefont {Pironio}, \citenamefont {Bancal},\
  and\ \citenamefont {Scarani}}]{pironio2011extremal}%
  \BibitemOpen
  \bibfield  {author} {\bibinfo {author} {\bibfnamefont {S.}~\bibnamefont
  {Pironio}}, \bibinfo {author} {\bibfnamefont {J.-D.}\ \bibnamefont {Bancal}},
  \ and\ \bibinfo {author} {\bibfnamefont {V.}~\bibnamefont {Scarani}},\
  }\href@noop {} {\bibfield  {journal} {\bibinfo  {journal} {Journal of Physics
  A: Mathematical and Theoretical}\ }\textbf {\bibinfo {volume} {44}},\
  \bibinfo {pages} {065303} (\bibinfo {year} {2011}{\natexlab{a}})}\BibitemShut
  {NoStop}%
\bibitem [{\citenamefont {Svetlichny}(1987)}]{PhysRevD.35.3066}%
  \BibitemOpen
  \bibfield  {author} {\bibinfo {author} {\bibfnamefont {G.}~\bibnamefont
  {Svetlichny}},\ }\href {\doibase 10.1103/PhysRevD.35.3066} {\bibfield
  {journal} {\bibinfo  {journal} {Phys. Rev. D}\ }\textbf {\bibinfo {volume}
  {35}},\ \bibinfo {pages} {3066} (\bibinfo {year} {1987})}\BibitemShut
  {NoStop}%
\bibitem [{\citenamefont {Seevinck}\ and\ \citenamefont
  {Svetlichny}(2002)}]{PhysRevLett.89.060401}%
  \BibitemOpen
  \bibfield  {author} {\bibinfo {author} {\bibfnamefont {M.}~\bibnamefont
  {Seevinck}}\ and\ \bibinfo {author} {\bibfnamefont {G.}~\bibnamefont
  {Svetlichny}},\ }\href {\doibase 10.1103/PhysRevLett.89.060401} {\bibfield
  {journal} {\bibinfo  {journal} {Phys. Rev. Lett.}\ }\textbf {\bibinfo
  {volume} {89}},\ \bibinfo {pages} {060401} (\bibinfo {year}
  {2002})}\BibitemShut {NoStop}%
\bibitem [{\citenamefont {Greenberger}\ \emph {et~al.}(1990)\citenamefont
  {Greenberger}, \citenamefont {Horne}, \citenamefont {Shimony},\ and\
  \citenamefont {Zeilinger}}]{greenberger1990bell}%
  \BibitemOpen
  \bibfield  {author} {\bibinfo {author} {\bibfnamefont {D.~M.}\ \bibnamefont
  {Greenberger}}, \bibinfo {author} {\bibfnamefont {M.~A.}\ \bibnamefont
  {Horne}}, \bibinfo {author} {\bibfnamefont {A.}~\bibnamefont {Shimony}}, \
  and\ \bibinfo {author} {\bibfnamefont {A.}~\bibnamefont {Zeilinger}},\
  }\href@noop {} {\bibfield  {journal} {\bibinfo  {journal} {American Journal
  of Physics}\ }\textbf {\bibinfo {volume} {58}},\ \bibinfo {pages} {1131}
  (\bibinfo {year} {1990})}\BibitemShut {NoStop}%
\bibitem [{\citenamefont {Pan}\ \emph {et~al.}(2000)\citenamefont {Pan},
  \citenamefont {Bouwmeester}, \citenamefont {Daniell}, \citenamefont
  {Weinfurter},\ and\ \citenamefont {Zeilinger}}]{pan2000experimental}%
  \BibitemOpen
  \bibfield  {author} {\bibinfo {author} {\bibfnamefont {J.-W.}\ \bibnamefont
  {Pan}}, \bibinfo {author} {\bibfnamefont {D.}~\bibnamefont {Bouwmeester}},
  \bibinfo {author} {\bibfnamefont {M.}~\bibnamefont {Daniell}}, \bibinfo
  {author} {\bibfnamefont {H.}~\bibnamefont {Weinfurter}}, \ and\ \bibinfo
  {author} {\bibfnamefont {A.}~\bibnamefont {Zeilinger}},\ }\href@noop {}
  {\bibfield  {journal} {\bibinfo  {journal} {Nature}\ }\textbf {\bibinfo
  {volume} {403}},\ \bibinfo {pages} {515} (\bibinfo {year}
  {2000})}\BibitemShut {NoStop}%
\bibitem [{\citenamefont {Grasselli}\ \emph {et~al.}(2018)\citenamefont
  {Grasselli}, \citenamefont {Kampermann},\ and\ \citenamefont
  {Bru{\SS}}}]{Grasselli2018}%
  \BibitemOpen
  \bibfield  {author} {\bibinfo {author} {\bibfnamefont {F.}~\bibnamefont
  {Grasselli}}, \bibinfo {author} {\bibfnamefont {H.}~\bibnamefont
  {Kampermann}}, \ and\ \bibinfo {author} {\bibfnamefont {D.}~\bibnamefont
  {Bru{\SS}}},\ }\href {\doibase 10.1088/1367-2630/aaec34} {\bibfield
  {journal} {\bibinfo  {journal} {New Journal of Physics}\ }\textbf {\bibinfo
  {volume} {20}},\ \bibinfo {pages} {113014} (\bibinfo {year}
  {2018})}\BibitemShut {NoStop}%
\bibitem [{\citenamefont {Grasselli}\ \emph {et~al.}(2019)\citenamefont
  {Grasselli}, \citenamefont {Kampermann},\ and\ \citenamefont
  {Bru{\SS}}}]{Grasselli2019}%
  \BibitemOpen
  \bibfield  {author} {\bibinfo {author} {\bibfnamefont {F.}~\bibnamefont
  {Grasselli}}, \bibinfo {author} {\bibfnamefont {H.}~\bibnamefont
  {Kampermann}}, \ and\ \bibinfo {author} {\bibfnamefont {D.}~\bibnamefont
  {Bru{\SS}}},\ }\href {\doibase 10.1088/1367-2630/ab573e} {\bibfield
  {journal} {\bibinfo  {journal} {New Journal of Physics}\ }\textbf {\bibinfo
  {volume} {21}},\ \bibinfo {pages} {123002} (\bibinfo {year}
  {2019})}\BibitemShut {NoStop}%
\bibitem [{\citenamefont {Fu}\ \emph {et~al.}(2015)\citenamefont {Fu},
  \citenamefont {Yin}, \citenamefont {Chen},\ and\ \citenamefont
  {Chen}}]{PhysRevLett.114.090501}%
  \BibitemOpen
  \bibfield  {author} {\bibinfo {author} {\bibfnamefont {Y.}~\bibnamefont
  {Fu}}, \bibinfo {author} {\bibfnamefont {H.-L.}\ \bibnamefont {Yin}},
  \bibinfo {author} {\bibfnamefont {T.-Y.}\ \bibnamefont {Chen}}, \ and\
  \bibinfo {author} {\bibfnamefont {Z.-B.}\ \bibnamefont {Chen}},\ }\href
  {\doibase 10.1103/PhysRevLett.114.090501} {\bibfield  {journal} {\bibinfo
  {journal} {Phys. Rev. Lett.}\ }\textbf {\bibinfo {volume} {114}},\ \bibinfo
  {pages} {090501} (\bibinfo {year} {2015})}\BibitemShut {NoStop}%
\bibitem [{\citenamefont {Proietti}\ \emph {et~al.}(2021)\citenamefont
  {Proietti}, \citenamefont {Ho}, \citenamefont {Grasselli}, \citenamefont
  {Barrow}, \citenamefont {Malik},\ and\ \citenamefont
  {Fedrizzi}}]{sciadv.abe0395}%
  \BibitemOpen
  \bibfield  {author} {\bibinfo {author} {\bibfnamefont {M.}~\bibnamefont
  {Proietti}}, \bibinfo {author} {\bibfnamefont {J.}~\bibnamefont {Ho}},
  \bibinfo {author} {\bibfnamefont {F.}~\bibnamefont {Grasselli}}, \bibinfo
  {author} {\bibfnamefont {P.}~\bibnamefont {Barrow}}, \bibinfo {author}
  {\bibfnamefont {M.}~\bibnamefont {Malik}}, \ and\ \bibinfo {author}
  {\bibfnamefont {A.}~\bibnamefont {Fedrizzi}},\ }\href {\doibase
  10.1126/sciadv.abe039} {\bibfield  {journal} {\bibinfo  {journal} {Science
  Advances}\ }\textbf {\bibinfo {volume} {7}},\ \bibinfo {pages} {abe0395}
  (\bibinfo {year} {2021})}\BibitemShut {NoStop}%
\bibitem [{\citenamefont {Ac\'{\i}n}\ \emph {et~al.}(2012)\citenamefont
  {Ac\'{\i}n}, \citenamefont {Massar},\ and\ \citenamefont
  {Pironio}}]{PhysRevLett.108.100402}%
  \BibitemOpen
  \bibfield  {author} {\bibinfo {author} {\bibfnamefont {A.}~\bibnamefont
  {Ac\'{\i}n}}, \bibinfo {author} {\bibfnamefont {S.}~\bibnamefont {Massar}}, \
  and\ \bibinfo {author} {\bibfnamefont {S.}~\bibnamefont {Pironio}},\ }\href
  {\doibase 10.1103/PhysRevLett.108.100402} {\bibfield  {journal} {\bibinfo
  {journal} {Phys. Rev. Lett.}\ }\textbf {\bibinfo {volume} {108}},\ \bibinfo
  {pages} {100402} (\bibinfo {year} {2012})}\BibitemShut {NoStop}%
\bibitem [{\citenamefont {Ribeiro}\ \emph {et~al.}(2018)\citenamefont
  {Ribeiro}, \citenamefont {Murta},\ and\ \citenamefont
  {Wehner}}]{PhysRevA.97.022307}%
  \BibitemOpen
  \bibfield  {author} {\bibinfo {author} {\bibfnamefont {J.}~\bibnamefont
  {Ribeiro}}, \bibinfo {author} {\bibfnamefont {G.}~\bibnamefont {Murta}}, \
  and\ \bibinfo {author} {\bibfnamefont {S.}~\bibnamefont {Wehner}},\ }\href
  {\doibase 10.1103/PhysRevA.97.022307} {\bibfield  {journal} {\bibinfo
  {journal} {Phys. Rev. A}\ }\textbf {\bibinfo {volume} {97}},\ \bibinfo
  {pages} {022307} (\bibinfo {year} {2018})}\BibitemShut {NoStop}%
\bibitem [{\citenamefont {Holz}\ \emph {et~al.}(2020)\citenamefont {Holz},
  \citenamefont {Kampermann},\ and\ \citenamefont
  {Bru\ss{}}}]{PhysRevResearch.2.023251}%
  \BibitemOpen
  \bibfield  {author} {\bibinfo {author} {\bibfnamefont {T.}~\bibnamefont
  {Holz}}, \bibinfo {author} {\bibfnamefont {H.}~\bibnamefont {Kampermann}}, \
  and\ \bibinfo {author} {\bibfnamefont {D.}~\bibnamefont {Bru\ss{}}},\ }\href
  {\doibase 10.1103/PhysRevResearch.2.023251} {\bibfield  {journal} {\bibinfo
  {journal} {Phys. Rev. Research}\ }\textbf {\bibinfo {volume} {2}},\ \bibinfo
  {pages} {023251} (\bibinfo {year} {2020})}\BibitemShut {NoStop}%
\bibitem [{\citenamefont {Pironio}\ \emph {et~al.}(2009)\citenamefont
  {Pironio}, \citenamefont {Acín}, \citenamefont {Brunner}, \citenamefont
  {Gisin}, \citenamefont {Massar},\ and\ \citenamefont
  {Scarani}}]{Pironio_2009}%
  \BibitemOpen
  \bibfield  {author} {\bibinfo {author} {\bibfnamefont {S.}~\bibnamefont
  {Pironio}}, \bibinfo {author} {\bibfnamefont {A.}~\bibnamefont {Acín}},
  \bibinfo {author} {\bibfnamefont {N.}~\bibnamefont {Brunner}}, \bibinfo
  {author} {\bibfnamefont {N.}~\bibnamefont {Gisin}}, \bibinfo {author}
  {\bibfnamefont {S.}~\bibnamefont {Massar}}, \ and\ \bibinfo {author}
  {\bibfnamefont {V.}~\bibnamefont {Scarani}},\ }\href {\doibase
  10.1088/1367-2630/11/4/045021} {\bibfield  {journal} {\bibinfo  {journal}
  {New Journal of Physics}\ }\textbf {\bibinfo {volume} {11}},\ \bibinfo
  {pages} {045021} (\bibinfo {year} {2009})}\BibitemShut {NoStop}%
\bibitem [{\citenamefont {Ac\'{\i}n}\ \emph {et~al.}(2007)\citenamefont
  {Ac\'{\i}n}, \citenamefont {Brunner}, \citenamefont {Gisin}, \citenamefont
  {Massar}, \citenamefont {Pironio},\ and\ \citenamefont
  {Scarani}}]{PhysRevLett.98.230501}%
  \BibitemOpen
  \bibfield  {author} {\bibinfo {author} {\bibfnamefont {A.}~\bibnamefont
  {Ac\'{\i}n}}, \bibinfo {author} {\bibfnamefont {N.}~\bibnamefont {Brunner}},
  \bibinfo {author} {\bibfnamefont {N.}~\bibnamefont {Gisin}}, \bibinfo
  {author} {\bibfnamefont {S.}~\bibnamefont {Massar}}, \bibinfo {author}
  {\bibfnamefont {S.}~\bibnamefont {Pironio}}, \ and\ \bibinfo {author}
  {\bibfnamefont {V.}~\bibnamefont {Scarani}},\ }\href {\doibase
  10.1103/PhysRevLett.98.230501} {\bibfield  {journal} {\bibinfo  {journal}
  {Phys. Rev. Lett.}\ }\textbf {\bibinfo {volume} {98}},\ \bibinfo {pages}
  {230501} (\bibinfo {year} {2007})}\BibitemShut {NoStop}%
\bibitem [{\citenamefont {Vazirani}\ and\ \citenamefont
  {Vidick}(2014)}]{PhysRevLett.113.140501}%
  \BibitemOpen
  \bibfield  {author} {\bibinfo {author} {\bibfnamefont {U.}~\bibnamefont
  {Vazirani}}\ and\ \bibinfo {author} {\bibfnamefont {T.}~\bibnamefont
  {Vidick}},\ }\href {\doibase 10.1103/PhysRevLett.113.140501} {\bibfield
  {journal} {\bibinfo  {journal} {Phys. Rev. Lett.}\ }\textbf {\bibinfo
  {volume} {113}},\ \bibinfo {pages} {140501} (\bibinfo {year}
  {2014})}\BibitemShut {NoStop}%
\bibitem [{\citenamefont {Woodhead}\ \emph {et~al.}(2021)\citenamefont
  {Woodhead}, \citenamefont {Ac{\'{i}}n},\ and\ \citenamefont
  {Pironio}}]{Woodhead2021deviceindependent}%
  \BibitemOpen
  \bibfield  {author} {\bibinfo {author} {\bibfnamefont {E.}~\bibnamefont
  {Woodhead}}, \bibinfo {author} {\bibfnamefont {A.}~\bibnamefont
  {Ac{\'{i}}n}}, \ and\ \bibinfo {author} {\bibfnamefont {S.}~\bibnamefont
  {Pironio}},\ }\href {\doibase 10.22331/q-2021-04-26-443} {\bibfield
  {journal} {\bibinfo  {journal} {{Quantum}}\ }\textbf {\bibinfo {volume}
  {5}},\ \bibinfo {pages} {443} (\bibinfo {year} {2021})}\BibitemShut {NoStop}%
\bibitem [{\citenamefont {Ho}\ \emph {et~al.}(2020)\citenamefont {Ho},
  \citenamefont {Sekatski}, \citenamefont {Tan}, \citenamefont {Renner},
  \citenamefont {Bancal},\ and\ \citenamefont
  {Sangouard}}]{PhysRevLett.124.230502}%
  \BibitemOpen
  \bibfield  {author} {\bibinfo {author} {\bibfnamefont {M.}~\bibnamefont
  {Ho}}, \bibinfo {author} {\bibfnamefont {P.}~\bibnamefont {Sekatski}},
  \bibinfo {author} {\bibfnamefont {E.~Y.-Z.}\ \bibnamefont {Tan}}, \bibinfo
  {author} {\bibfnamefont {R.}~\bibnamefont {Renner}}, \bibinfo {author}
  {\bibfnamefont {J.-D.}\ \bibnamefont {Bancal}}, \ and\ \bibinfo {author}
  {\bibfnamefont {N.}~\bibnamefont {Sangouard}},\ }\href {\doibase
  10.1103/PhysRevLett.124.230502} {\bibfield  {journal} {\bibinfo  {journal}
  {Phys. Rev. Lett.}\ }\textbf {\bibinfo {volume} {124}},\ \bibinfo {pages}
  {230502} (\bibinfo {year} {2020})}\BibitemShut {NoStop}%
\bibitem [{\citenamefont {Sekatski}\ \emph {et~al.}(2021)\citenamefont
  {Sekatski}, \citenamefont {Bancal}, \citenamefont {Valcarce}, \citenamefont
  {Tan}, \citenamefont {Renner},\ and\ \citenamefont
  {Sangouard}}]{Sekatski2021deviceindependent}%
  \BibitemOpen
  \bibfield  {author} {\bibinfo {author} {\bibfnamefont {P.}~\bibnamefont
  {Sekatski}}, \bibinfo {author} {\bibfnamefont {J.-D.}\ \bibnamefont
  {Bancal}}, \bibinfo {author} {\bibfnamefont {X.}~\bibnamefont {Valcarce}},
  \bibinfo {author} {\bibfnamefont {E.~Y.-Z.}\ \bibnamefont {Tan}}, \bibinfo
  {author} {\bibfnamefont {R.}~\bibnamefont {Renner}}, \ and\ \bibinfo {author}
  {\bibfnamefont {N.}~\bibnamefont {Sangouard}},\ }\href {\doibase
  10.22331/q-2021-04-26-444} {\bibfield  {journal} {\bibinfo  {journal}
  {{Quantum}}\ }\textbf {\bibinfo {volume} {5}},\ \bibinfo {pages} {444}
  (\bibinfo {year} {2021})}\BibitemShut {NoStop}%
\bibitem [{\citenamefont {Xu}\ \emph {et~al.}(2022)\citenamefont {Xu},
  \citenamefont {Zhang}, \citenamefont {Zhang},\ and\ \citenamefont
  {Pan}}]{PhysRevLett.128.110506}%
  \BibitemOpen
  \bibfield  {author} {\bibinfo {author} {\bibfnamefont {F.}~\bibnamefont
  {Xu}}, \bibinfo {author} {\bibfnamefont {Y.-Z.}\ \bibnamefont {Zhang}},
  \bibinfo {author} {\bibfnamefont {Q.}~\bibnamefont {Zhang}}, \ and\ \bibinfo
  {author} {\bibfnamefont {J.-W.}\ \bibnamefont {Pan}},\ }\href {\doibase
  10.1103/PhysRevLett.128.110506} {\bibfield  {journal} {\bibinfo  {journal}
  {Phys. Rev. Lett.}\ }\textbf {\bibinfo {volume} {128}},\ \bibinfo {pages}
  {110506} (\bibinfo {year} {2022})}\BibitemShut {NoStop}%
\bibitem [{\citenamefont {Gonzales-Ureta}\ \emph {et~al.}(2021)\citenamefont
  {Gonzales-Ureta}, \citenamefont {Predojevi\ifmmode~\acute{c}\else
  \'{c}\fi{}},\ and\ \citenamefont {Cabello}}]{PhysRevA.103.052436}%
  \BibitemOpen
  \bibfield  {author} {\bibinfo {author} {\bibfnamefont {J.~R.}\ \bibnamefont
  {Gonzales-Ureta}}, \bibinfo {author} {\bibfnamefont {A.}~\bibnamefont
  {Predojevi\ifmmode~\acute{c}\else \'{c}\fi{}}}, \ and\ \bibinfo {author}
  {\bibfnamefont {A.}~\bibnamefont {Cabello}},\ }\href {\doibase
  10.1103/PhysRevA.103.052436} {\bibfield  {journal} {\bibinfo  {journal}
  {Phys. Rev. A}\ }\textbf {\bibinfo {volume} {103}},\ \bibinfo {pages}
  {052436} (\bibinfo {year} {2021})}\BibitemShut {NoStop}%
\bibitem [{\citenamefont {Acín}\ \emph {et~al.}(2006)\citenamefont {Acín},
  \citenamefont {Massar},\ and\ \citenamefont {Pironio}}]{Acin2006}%
  \BibitemOpen
  \bibfield  {author} {\bibinfo {author} {\bibfnamefont {A.}~\bibnamefont
  {Acín}}, \bibinfo {author} {\bibfnamefont {S.}~\bibnamefont {Massar}}, \
  and\ \bibinfo {author} {\bibfnamefont {S.}~\bibnamefont {Pironio}},\ }\href
  {\doibase 10.1088/1367-2630/8/8/126} {\bibfield  {journal} {\bibinfo
  {journal} {New Journal of Physics}\ }\textbf {\bibinfo {volume} {8}},\
  \bibinfo {pages} {126} (\bibinfo {year} {2006})}\BibitemShut {NoStop}%
\bibitem [{\citenamefont {Ac\'{\i}n}\ \emph {et~al.}(2006)\citenamefont
  {Ac\'{\i}n}, \citenamefont {Gisin},\ and\ \citenamefont
  {Masanes}}]{PhysRevLett.97.120405}%
  \BibitemOpen
  \bibfield  {author} {\bibinfo {author} {\bibfnamefont {A.}~\bibnamefont
  {Ac\'{\i}n}}, \bibinfo {author} {\bibfnamefont {N.}~\bibnamefont {Gisin}}, \
  and\ \bibinfo {author} {\bibfnamefont {L.}~\bibnamefont {Masanes}},\ }\href
  {\doibase 10.1103/PhysRevLett.97.120405} {\bibfield  {journal} {\bibinfo
  {journal} {Phys. Rev. Lett.}\ }\textbf {\bibinfo {volume} {97}},\ \bibinfo
  {pages} {120405} (\bibinfo {year} {2006})}\BibitemShut {NoStop}%
\bibitem [{\citenamefont {Farkas}\ \emph {et~al.}(2021)\citenamefont {Farkas},
  \citenamefont {Balanz\'o-Juand\'o}, \citenamefont {\L{}ukanowski},
  \citenamefont {Ko\l{}ody\ifmmode~\acute{n}\else \'{n}\fi{}ski},\ and\
  \citenamefont {Ac\'{\i}n}}]{PhysRevLett.127.050503}%
  \BibitemOpen
  \bibfield  {author} {\bibinfo {author} {\bibfnamefont {M.}~\bibnamefont
  {Farkas}}, \bibinfo {author} {\bibfnamefont {M.}~\bibnamefont
  {Balanz\'o-Juand\'o}}, \bibinfo {author} {\bibfnamefont {K.}~\bibnamefont
  {\L{}ukanowski}}, \bibinfo {author} {\bibfnamefont {J.}~\bibnamefont
  {Ko\l{}ody\ifmmode~\acute{n}\else \'{n}\fi{}ski}}, \ and\ \bibinfo {author}
  {\bibfnamefont {A.}~\bibnamefont {Ac\'{\i}n}},\ }\href {\doibase
  10.1103/PhysRevLett.127.050503} {\bibfield  {journal} {\bibinfo  {journal}
  {Phys. Rev. Lett.}\ }\textbf {\bibinfo {volume} {127}},\ \bibinfo {pages}
  {050503} (\bibinfo {year} {2021})}\BibitemShut {NoStop}%
\bibitem [{\citenamefont {Werner}(1989)}]{PhysRevA.40.4277}%
  \BibitemOpen
  \bibfield  {author} {\bibinfo {author} {\bibfnamefont {R.~F.}\ \bibnamefont
  {Werner}},\ }\href {\doibase 10.1103/PhysRevA.40.4277} {\bibfield  {journal}
  {\bibinfo  {journal} {Phys. Rev. A}\ }\textbf {\bibinfo {volume} {40}},\
  \bibinfo {pages} {4277} (\bibinfo {year} {1989})}\BibitemShut {NoStop}%
\bibitem [{\citenamefont {Devetak}\ and\ \citenamefont
  {Winter}(2005)}]{Devetak2005}%
  \BibitemOpen
  \bibfield  {author} {\bibinfo {author} {\bibfnamefont {I.}~\bibnamefont
  {Devetak}}\ and\ \bibinfo {author} {\bibfnamefont {A.}~\bibnamefont
  {Winter}},\ }\href {\doibase https://doi.org/10.1098/rspa.2004.1372}
  {\bibfield  {journal} {\bibinfo  {journal} {Proc. R. Soc. A}\ }\textbf
  {\bibinfo {volume} {461}},\ \bibinfo {pages} {207} (\bibinfo {year}
  {2005})}\BibitemShut {NoStop}%
\bibitem [{\citenamefont {Pironio}\ \emph
  {et~al.}(2011{\natexlab{b}})\citenamefont {Pironio}, \citenamefont {Bancal},\
  and\ \citenamefont {Scarani}}]{Pironio2011}%
  \BibitemOpen
  \bibfield  {author} {\bibinfo {author} {\bibfnamefont {S.}~\bibnamefont
  {Pironio}}, \bibinfo {author} {\bibfnamefont {J.-D.}\ \bibnamefont {Bancal}},
  \ and\ \bibinfo {author} {\bibfnamefont {V.}~\bibnamefont {Scarani}},\ }\href
  {\doibase 10.1088/1751-8113/44/6/065303} {\bibfield  {journal} {\bibinfo
  {journal} {Journal of Physics A: Mathematical and Theoretical}\ }\textbf
  {\bibinfo {volume} {44}},\ \bibinfo {pages} {065303} (\bibinfo {year}
  {2011}{\natexlab{b}})}\BibitemShut {NoStop}%
\end{thebibliography}%

\end{document}